\documentclass[journal]{IEEEtran}
\ifCLASSINFOpdf
\else
\fi
\usepackage{graphicx}
\usepackage{subcaption}
\usepackage{algorithm,algorithmic}
\usepackage{amsmath}
\usepackage{multirow}
\usepackage{breqn}
\usepackage[table]{xcolor}
\definecolor{skyblue}{rgb}{0.53, 0.81, 0.92}
\definecolor{flax}{rgb}{0.93, 0.86, 0.51}
\newcommand{\norm}[1]{\left\lVert#1\right\rVert}
\begin{document}
%
\title{DNN Transfer Learning from Diversified Micro-Doppler for Motion Classification}
%
%
%

\author{Mehmet Saygin Seyfioglu, ~\IEEEmembership{Student Member,~IEEE,} 
Baris Erol, ~\IEEEmembership{Student Member,~IEEE,} 
Sevgi Zubeyde Gurbuz, ~\IEEEmembership{Senior Member,~IEEE,} 
Moeness G. Amin, ~\IEEEmembership{Fellow,~IEEE,}

\thanks{M.S. Seyfioglu is with the Department of Electrical and Electronics Engineering, TOBB University of Economics and Technology, Ankara, Turkey, email:  msseyfioglu@etu.edu.tr.}
\thanks{B. Erol and M.G. Amin are with the Center for Advanced Communications, Villanova University, PA, email: berol@villanova.edu, moeness.amin@villanova.edu}
\thanks{S.Z. Gurbuz is with the Department
of Electrical and Computer Engineering, University of Alabama, Tuscaloosa,
AL, 30332 USA e-mail: szgurbuz@ua.edu.}

\thanks{Manuscript submitted January 12, 2018.}}

%
%

\markboth{ Submitted to IEEE Transactions on Aerospace and Electronic Systems}%
{Shell \MakeLowercase{\textit{et al.}}: Bare Demo of IEEEtran.cls for IEEE Journals}
%



\maketitle

\begin{abstract}
Recently, deep neural networks (DNNs) have been the subject of intense research for the classification of radio frequency (RF) signals, such as synthetic aperture radar (SAR) imagery or micro-Doppler signatures. However, a fundamental challenge is the typically small amount of data available due to the high costs and resources required for measurements.  Small datasets limit the depth of DNNs implementable, and limit performance.  In this work, a novel method for generating diversified radar micro-Doppler signatures using Kinect-based motion capture simulations is proposed as a training database for transfer learning with DNNs.  In particular, it is shown that together with residual learning, the proposed DivNet approach allows for the construction of deeper neural networks and offers improved performance in comparison to transfer learning from optical imagery.  Furthermore, it is shown that initializing the network using diversified synthetic micro-Doppler signatures enables not only robust performance for previously unseen target profiles, but also class generalization. Results are presented for 7-class and 11-class human activity recognition scenarios using a 4-GHz continuous wave (CW) software-defined radar.  
\end{abstract}


%
\IEEEpeerreviewmaketitle

\section{Introduction}
%
%
%
%
\IEEEPARstart{D}{eep} neural networks (DNNs)  have recently attracted great interest in the radar community as a means for learning fine representations of the underlying data in a variety of applications, such as static object recognition \cite{lombacher16}, synthetic aperture radar (SAR) target classification \cite{schwegmann17}-\nocite{wagner17}\nocite{li17}\nocite{lin17}\cite{chen16}, change detection \cite{geng17} \cite{gong16}, and image registration \cite{quan16}, as well as airborne phased array radar mode recognition \cite{li16}, and radar waveform recognition \cite{wang17}. DNNs have also been exploited in micro-Doppler based automatic target recognition studies relating to recognition of drones \cite{mendis2016deep} \cite{kim17}, human activities \cite{kim2016human}-\nocite{trommel2016multi}\nocite{jokanovic2016radar}\nocite{seyfioglu17}\cite{kwon17}, and hand gestures \cite{molchanov15hand} \cite{kim2016hand}. 

A common challenge in all these applications is that the datasets available are typically quite limited, often on the order of hundreds of samples for micro-Doppler, while reaching levels of thousands for SAR imagery, as shown in Table \ref{t1}. One approach for dealing with low sample support is to use sparsely connected layers, as opposed to fully connected layers, to reduce the total number of parameters in the network \cite{Chen16sparse}. Convolutional autoencoders (CAE) have been proposed \cite{SayginCAE} as a DNN architecture that uses unsupervised pre-training to mitigate measured data requirements.  However, when less than 550 training samples are available, CAEs have been outperformed \cite{SayginGSRL2017} by transfer learning from other domains, such as optical imagery.  Another approach to preventing overfitting is to limit the depth of the DNNs; however, this also limits the ability of the DNN to accurately model complex data representations and, in turn, the classification accuracy achievable.  Another disadvantage of small training datasets is the network's inability to generalize to previously unobserved targets: i.e., classification accuracy plummets if the new person observed has a significant difference in gait speed, height, body structure or walking style than the subjects used to form the training database.  As a result, training data should strive to span possible and broad target signature variations for robust performance \cite{Ding16}.

One way to both increase the amount of training data as well as in-class variations is to use simulated data.  In computer vision, Generative Adversarial Networks (GANs) \cite{GANSgoodfellow} have been proposed as a means to generate highly realistic simulated images \cite{ShrivastavaGAN}.  However, one difficulty in training GANs is the potential for the generative network to collapse and map all outputs to the same data \cite{Tarik}.  Moreover, the phenomenology of radar data differs significantly from that of optical imagery.  This difference is attributed to the fact that radar measurements of human motion are fundamentally a continuous stream of non-stationary signals. Using spectrograms, these signals are expressed in the time-frequency domain and, as such, can be presented as an image to the DNN - a snapshot of one fixed window in time.  But the patterns in this image of the time-frequency representation have deeper meanings that relate to the kinematics of the target and physics of electromagnetic signals.  Image-based variations created with GANs can thus be unrelated to the differences in micro-Doppler signature caused by physical variations observed in the gait of different people, thereby generating misleading or outright erroneous training data.

As a result, physical model-based approaches have been preferred to simulate radar data.  The use of simulated signatures to classify real micro-Doppler measurements was first proposed in 2015 \cite{karabacak16}, where video motion capture (MOCAP) data made available from the Carnegie Mellon University, Graphics Lab  Motion Capture Database \cite{CMUlib} was used to classify measured data with only a 1\% difference in performance as when only real data was used for training.  Simulations have also been used to generate training data for classification of SAR images \cite{odegaard16}, as well as high resolution range profiles \cite{lunden16}.  More recently, kinematic models \cite{Khomchuk16}, or MOCAP data from low-cost devices, such as Kinect \cite{Erol_Kinect15}, have been utilized to simulate human radar returns in applications such as human activity recognition \cite{blasch2005}-\nocite{ram2008}\nocite{ram2010}\nocite{Gurbuz15}\nocite{Eeden15}\cite{Tekeli16} and fall detection \cite{saho17}.  

However, because these MOCAP simulations are tied to obtaining real infrared and optical sensor data from test subjects, as with radar data, the size of the dataset is still limited by the human effort, time and cost of data collections. In contrast, this paper demonstrates that just a small sub-set of MOCAP data can be used to generate a large number of simulated micro-Doppler signatures that spans the range of human body sizes and speeds, while also emulating individualized gait variations. This is accomplished by applying transformations on the underlying skeletal structure tracked by Kinect, such as scaling the skeletal dimensions to model different body sizes, scaling the time dimension to model different motion speeds, and perturbing the parameters of a parametric model of time-varying joint positions to model individualized gait style. Although there is no guarantee that every generated signature is fully compatible with the kinematic constraints of human motion, the extent of the discrepancy is much more limited by the underlying skeletal model in comparison to GANs and compensated for by using transfer learning - the diversified signatures are used only as a source for initializing the DNN.  A second stage of training on measured signatures is still performed to fine-tune network parameters, but due to the initial pre-training, only a minimal amount of measured data is required.

\begin{table}[!t]
\centering
\caption{DNN training data size used in recent works.}
\begin{tabular}{|c|c|c|c|}
\hline
\textbf{Application} & \textbf{Reference} & \textbf{\#  Classes} & \textbf{\# Training Data} \\
\hline\hline
human $\mu$D  & Jokanovic \cite{jokanovic2016radar} & 4 & 60 \\
\hline
drone $\mu$D  & Mendis \cite{mendis2016deep} & 3 & 210 \\
\hline
vehicular radar & Parashar \cite{parashar17} &  4 & 300 \\
\hline
hand gesture & Kim \cite{kim2016hand} & 10 & 450 \\
\hline
human $\mu$D & Kim  \cite{kim2016human} & 4 & 756 \\
\hline
human $\mu$D & Seyfioglu \cite{seyfioglu17} & 12 & 864 \\
\hline
vehicular radar & Lombacher \cite{lombacher16} & 10 & 3,397 \\
\hline
SAR  & Lin \cite{lin17} & 10 & 4,426 \\
\hline
SAR  & Chen\cite{chen16} & 10 & 5,173 \\
\hline
SAR  & Schwegmann \cite{schwegmann17} & 3 & 6,384 \\
\hline
SAR  & Wagner \cite{wagner17} & 10 & 6,874 \\
\hline
human $\mu$D & Trommel \cite{trommel2016multi} & 6 & 17,580 \\
\hline
drone $\mu$D & Kim \cite{kim17}  & 5 & 60,000 \\
\hline
SAR  & Li  \cite{li17} & 4 & 66,120 \\
\hline
\end{tabular}
\label{t1}
\end{table}

Transfer learning has been proposed for applications in SAR image classification \cite{kang16}, and moving target recognition \cite{yang16}.  More recently, Park, et al. \cite{park16} used transfer learning from VGGnet \cite{vggnet}, a 16-layer convolutional neural network (CNN), pre-trained with the ImageNet dataset \cite{imagenet} comprised of 1.5 million RGB images, to classify 5 different types of swimming. A classification accuracy of 80.3\% was achieved with fine-tuning on 756 micro-Doppler measurements.  Our results show, however, that network initialization using pertinent motion data is more effective than conventional random initialization or pre-training on unrelated optical imagery.  

Specific contributions of the proposed method include 

1) development of a model-based approach that exploits transformations of video motion capture data to generate arbitrarily large radio frequency (RF) micro-Doppler training datasets, 

2) improved classification accuracy as compared to transfer learning from optical imagery or training with measured data only,

3) improved target generalization performance through application of transformations on the underlying skeleton to generate training data that spans the wide range of probable target signatures, 

4) effective accuracy even under class generalization, 

5) a significant (10 to 20 fold) decrease in computational complexity relative to other networks proposed with transfer learning, e.g. VGGnet and ResNet-50, 

6) a reduction in the amount of required measured data, and 

7) increase in the depth of DNNs that can be designed for micro-Doppler classification.  

Using residual learning \cite{He16_ResNet} to prevent degradation in training accuracy as depth increases, the proposed  15 convolutional-layer residual neural network, DivNet-15, is shown to achieve an accuracy of 98\% for 7-classes, and 96\% when generalizing to 11-classes.

In Section II, the methodology for and validation of the diversified human micro-Doppler simulations, which exploit video motion capture, is described in detail.  In Section III, the measured data used as the test set is presented.  In Section IV, the design of CNN's are discussed in consideration of input dimensionality, depth, the number of neurons per layer and convolutional filter size. Section V presents a comparison of specific DNNs, based upon insights gain in Section IV.  In Section VI, the performance of the proposed architecture is contrasted with that of a randomly initialized CNN and transfer learning with VGGnet and ResNet. Classification accuracy, generalization, the impact of noise, as well as sensitivity of results to time shifting and dwell time are discussed in detail.  Feature visualizations are used to show that the generalized shapes learned by VGGnet are not as effective in representing the structure of micro-Doppler signatures as compared with the proposed simulation-based transfer learning approach.  Finally, in Section VII, key conclusions are presented.

\section{Simulated Radar Micro-Doppler Database}

\begin{figure*}[t!]
	\centering
	\begin{subfigure}[t]{.32\textwidth}
		\includegraphics [width=2.5in] {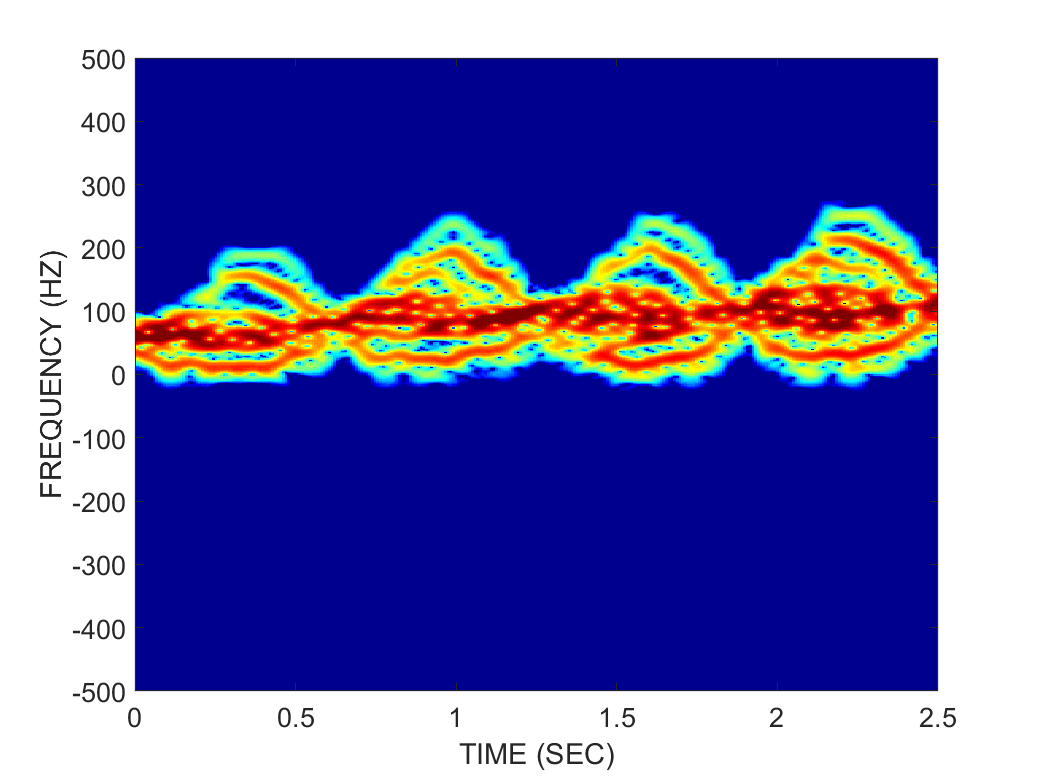}
		\caption{Walking}
	\end{subfigure}
	\begin{subfigure}[t]{.32\textwidth}
		\includegraphics [width=2.5in] {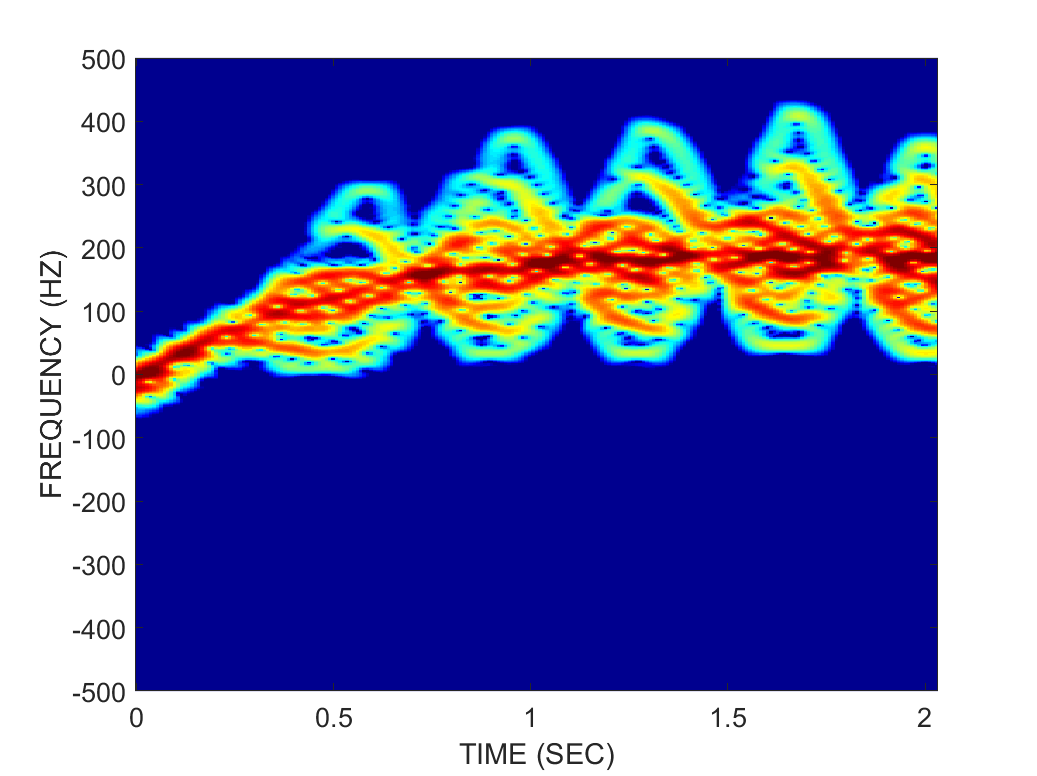}
		\caption{Running}
	\end{subfigure}
	\begin{subfigure}[t]{.32\textwidth}
		\includegraphics [width=2.5in] {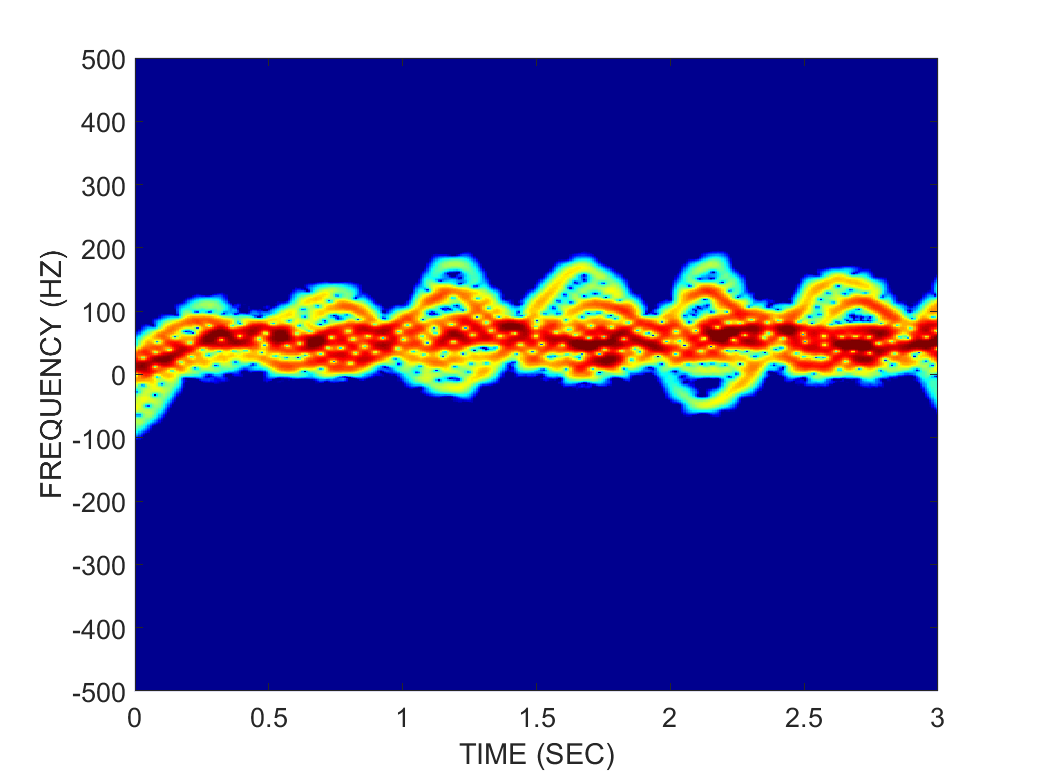}
		\caption{Limping}
	\end{subfigure}
	\begin{subfigure}[t]{.24\textwidth}
		\includegraphics [width=1.9in] {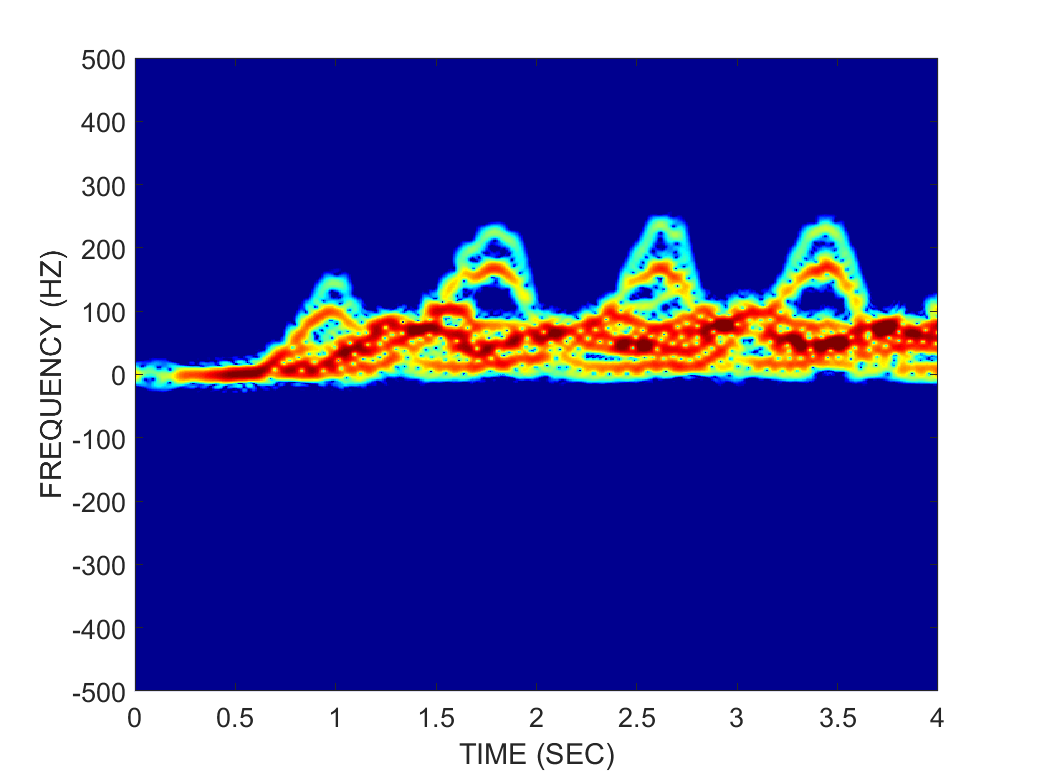}
		\caption{Cane}
	\end{subfigure}
	\centering
	\begin{subfigure}[t]{.24\textwidth}
		\includegraphics [width=1.9in] {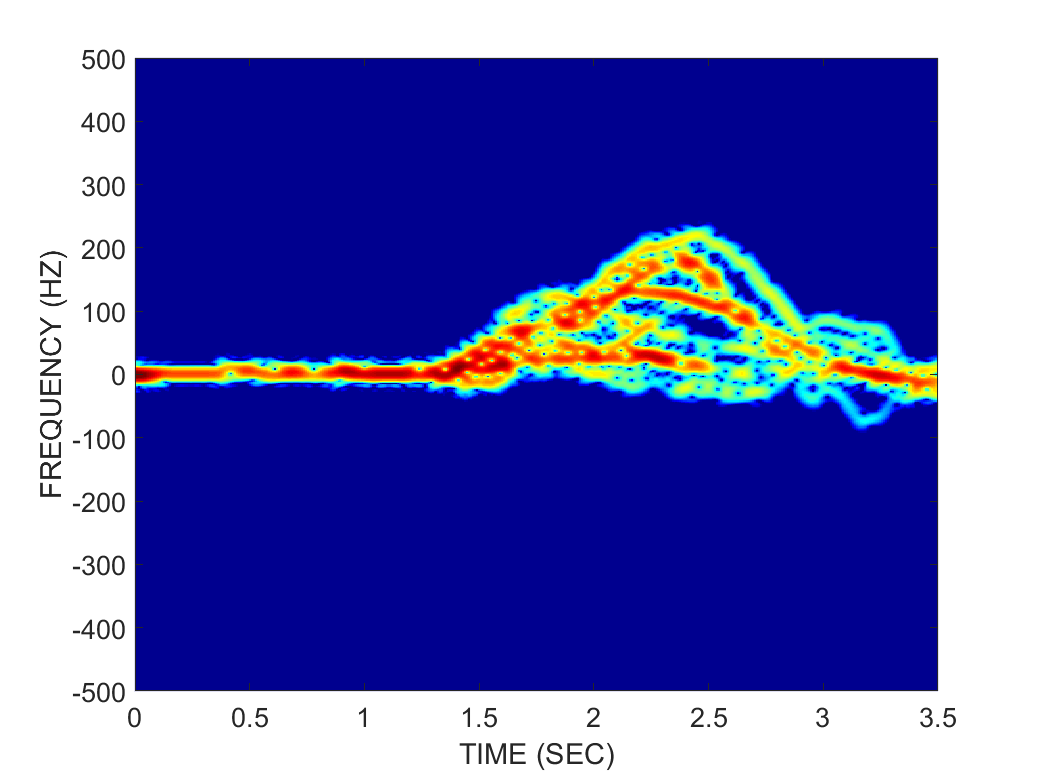}
		\caption{Falling}
	\end{subfigure}
	\begin{subfigure}[t]{.24\textwidth}
		\includegraphics [width=1.9in] {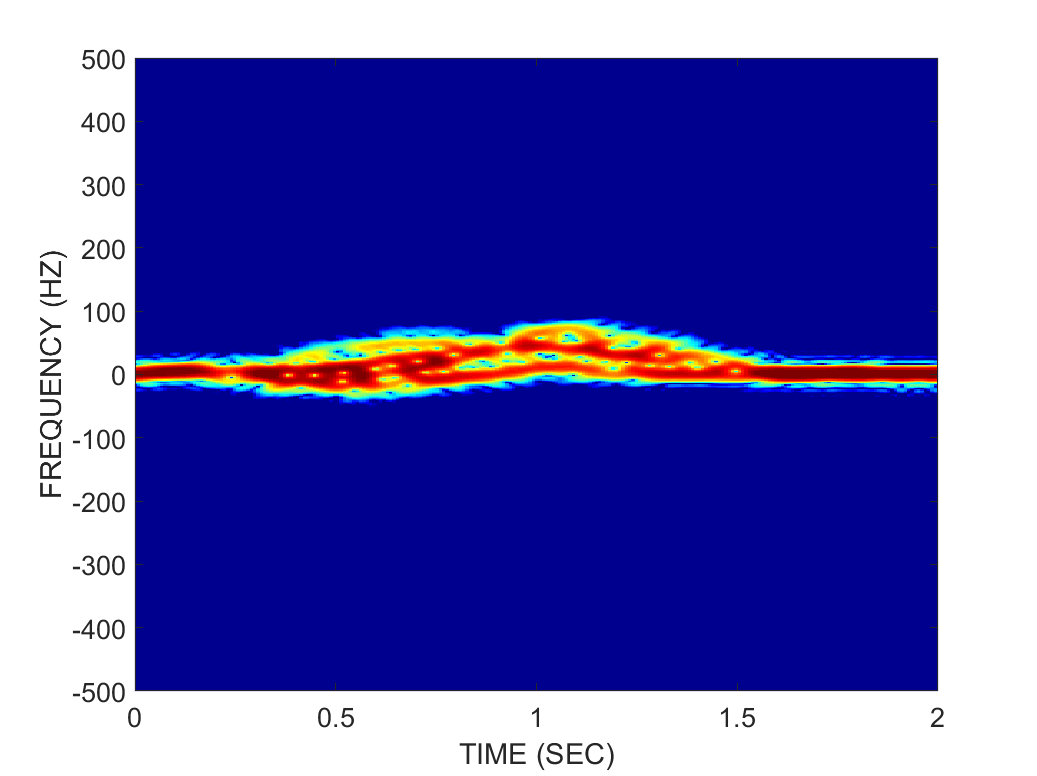}
		\caption{Sitting}
	\end{subfigure}
    	\begin{subfigure}[t]{.24\textwidth}
		\includegraphics [width=1.9in] {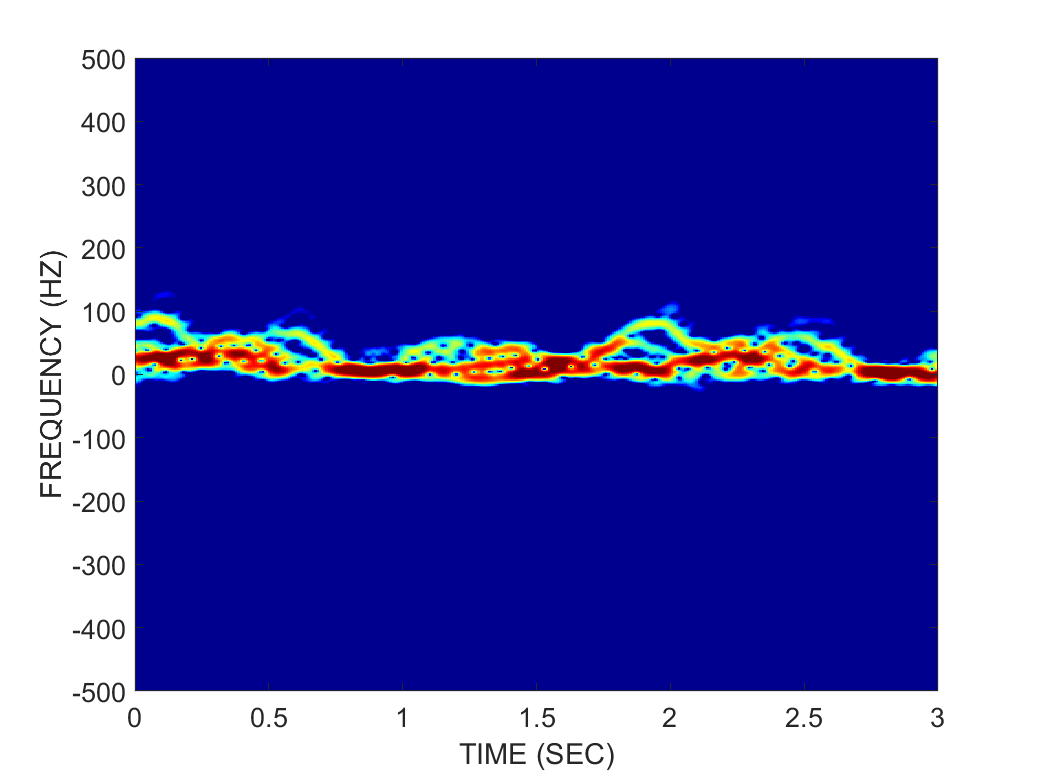}
		\caption{Walker}
	\end{subfigure}
	\caption{Kinect-based micro-Doppler signatures for a 15 GHz CW radar}\label{label-b}
\end{figure*}

There are two main approaches for simulating human micro-Doppler signatures \cite{IndoorRadar17}: kinematic modeling and MOCAP-based animation.  Both methods revolve around the idea of decomposing the human body into a finite number of parts modeled as point targets, and summing the radar returns \cite{vanDorp}, as modeled from the radar range equation.  The most widely used kinematic model in radar literature is the Boulic-Thalmann model \cite{boulic}, which is based on experimental studies of gait analysis and uses equations and charts to represent the time-varying motion of 17 different joints on the human body. The main disadvantage of the Boulic model is that it can be only applied to model walking. Generating full-body models for other types of non-rhythmic and aided periodic motions (e.g. walking with a cane and falling) still remains as a challenge.  Thus, more recently, MOCAP-based animations have gained in prevalence for micro-Doppler simulations.

\subsection{Kinect-Based Micro-Doppler Simulator}
In this work, the Kinect sensor is used as a markerless system for capturing the time-varying coordinate information of human joints needed for simulation of human micro-Doppler signatures \cite{Erol_Kinect15}.  First, the radar return from the human body is represented as the sum of reflected signals from a finite number ($K$) of point targets representing various body parts.   A total of 20 points were defined.  The Kinect measurements at these points are used in lieu of the time-varying range measurements typically obtained from radar.  Mathematically, the return signal from $K$ point targets for a continuous wave (CW) radar is
\begin{equation}
s_{h}(t)=\sum_{i=1}^{K} a_{t,i} e^{-j[(2\pi f_0)t + \frac{4 \pi}{\lambda} R_{t,i} ]}
\end{equation}
where $f_{0}$ is the transmit frequency, $\lambda$ is the wavelength, $t$ is time, $R_{t,i}$ is the time-varying range of each point target, as captured from the Kinect sensor, and $a_{t,i}$ is the amplitude as computed from the radar range equation
\begin{equation}
a_{t,i} = \frac{G \lambda \sqrt{P_{i} \sigma_{i}} } 
{(4 \pi)^{1.5} R_{t,i}^{2} \sqrt{L_{s}} \sqrt{L_{a}} }.
\end{equation}
Here, $G$ is the antenna gain, $P_{i}$ is the transmitter power, $\sigma_{i}$ is the radar cross section (RCS) for each point target, and $L_{s}$ and $L_{a}$ represent the system and atmospheric losses, respectively. 

Once the required ranges are estimated from the Kinect data, (1) can be computed for any human activity or
radar parameters, such as center frequency, bandwidth, and sampling frequency. Finally, the simulated micro-Doppler signatures are computed as the spectrogram ($\mathrm{S}$) -- modulus squared of short-time Fourier Transform (STFT) -- of the radar return:
\begin{equation}
\mathrm{S} = | STFT(n,\omega) |^{2} = \Bigg|\sum_{m=-\infty}^{\infty} 
\mathrm{s}[n+m] \mathrm{w}[m] \mathrm{e}^{-j \omega m}\Bigg|^2
\end{equation}
where $n$ is discretized time, and $\mathrm{w}[m]$ is a window function.  In this work, the data was sampled at 2.4 kHz, while a Hanning
window of length 256 and 128 overlap samples with 1024 total frequency points is applied in computing the spectrogram.  Simulated spectrograms were generated at 15 GHz and cropped to generate images that showed the internal structure of the micro-Doppler as clearly as possible for the initialization of the DivNet. Figure 1 shows the resulting simulated micro-Doppler signatures for seven different activity classes. It is important to note that in this paper we only consider spectrograms as a representative of quadratic time-frequency distributions (QTFDs). High resolution QTFDs or members of Cohen's class can be used to generate the input images for DNN, but their consideration is outside the scope of this paper.

\subsection{Diversification Methodology for $\mu$D Signatures}

In the Kinect-based radar micro-Doppler simulator, the 3-D coordinate measurements of 17 joints acquired from the Kinect sensor are used.  By changing this coordinate information it is possible to form a large activity database with sufficient intra and inter class variations that approximately emulates the diversity of human signatures caused by differences in height, speed and individual gait.

\subsubsection{Height and Speed Modifications}
The Kinect-based radar simulator permits modification of subject height and speed by scaling the time-varying joint position data along different axes.  For example, scaling along the z-axis, while keeping the x and y axes unchanged, modifies the subject height which results in different dimensions of body parts. These changes influence the RCS computations of individual point scatterers, thus also affecting the received signal power. The radar range equation defined in (2), requires the RCS of the individual body parts to compute the radar backscattering. For a human target, the RCS of each body part is represented by that of a sphere for the head, and ellipsoid for the torso and limbs. More specifically, the lower leg is defined as an ellipsoid and the corresponding RCS can be computed using radii along each dimension, roll angle and the direction angle to the receiving radar \cite{trott_stationary_2007}. Figure 2 presents the aspect angle dependencies of lower legs simulated with 3 different dimensions. The RCS of a short lower leg provides a more dispersed reflection along different aspect angles with a lower power level. On the other hand, the RCS of a long lower leg gives a strong reflection around -5 dBm in the case of a zero aspect angle. Therefore, in micro-Doppler signatures it is expected to observe an increased power level for taller subjects. 


Due to the fact that scaling is performed only along the z-axis, the stride rate, speed, and style of the motion remain unmodified in the first step of the diversification methodology. However, in real scenarios, a tall person typically walks or runs faster than a  shorter person, when all other body structure factors assumed equal, such as body mass, flexibility, and proportionality. This argument can be also proved kinematically by using Boulic-Thalmann walking model \cite{boulic}. In this model, human motion is defined using kinematic parameters such as, linear position, linear velocity, linear acceleration, angular position, angular velocity, and angular acceleration. Moreover, consistent with the previous argument between height and speed, the velocity of walking, $v$, in the model is defined in terms of thigh height as 

\begin{figure}[t]
 	\centering
	\includegraphics[width=3in]{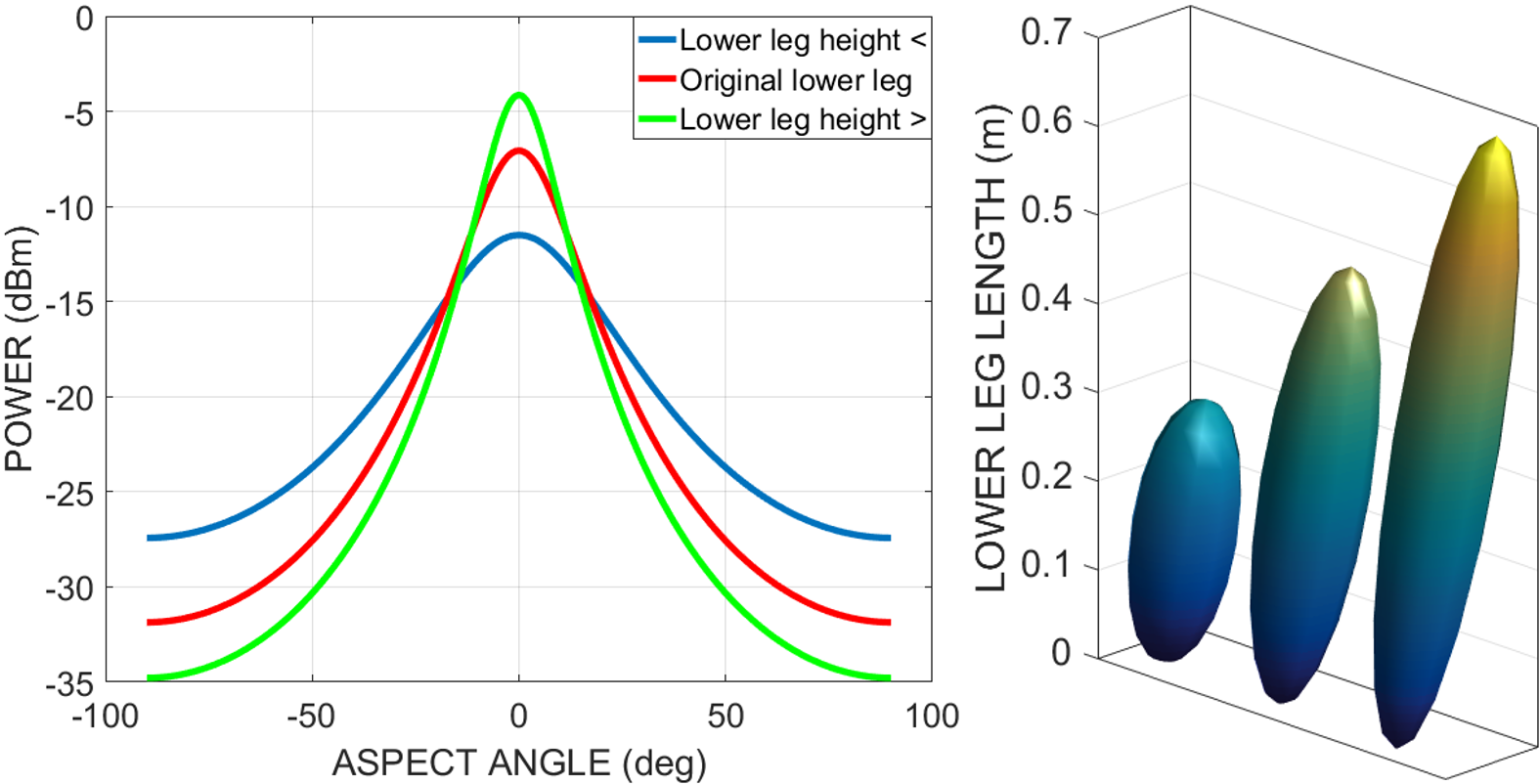}
	\caption{RCS angle dependency of lower legs with different dimensions}
	\label{fig:validacc30db4layers}
\end{figure}

\begin{equation}
\begin{aligned}
v = v_r \ H_{thigh} 
\end{aligned}
\end{equation}
where $v_r$ is the relative velocity in units of "thigh height" per second. By extending this expression, individual body part velocities can be computed. Assume that $\tilde{\mathrm{X}}_i(\tau.v_r)$ denotes the the position of a joint as a function of relative time $\tau$, and relative velocity, then the velocity of that individual body part can be defined as

\begin{equation}
\tilde{v}_i(\tau,v_r) = H_{thigh} \ \frac{v_r}{l_c} \ \frac{\partial }{\partial \tau} \tilde{\mathrm{X}}_i(\tau.v_r)
\end{equation}
where $l_c$ is the cycle length. Because of the relationship between height and speed, the x-axis distance position data is also scaled by a small amount relative to the change in height to alter the speed of the subject. However, this kinematic relation between leg length and average velocity is only proved for periodic motions, such as walking, and running. At this point another analysis is required for non-rhythmic motions such as, falling and sitting. For falling, upper body (head, shoulders, spine) and lower body (knee, calf) contain the most pronounced information in micro-Doppler signatures. A falling body moves with an increasing speed while it falls, and suddenly drops when it hits the ground, resulting in a tornado touching down shape. Sometimes, elderly people try to mitigate or slow down the fall by holding on the nearby objects which might extend the falling period. Apart from these situations, falling motion can be directly parameterized in a linear representation related to subject's height \cite{ModelDoppler}. Consider a simple example of falling rod from a vertical position.  Although, in radar data, the effect of the human upper body is much more complex than the motion of a rod, this comparison provides an important insight about the effect of the height. When the rod falls down its potential energy decreases and this would increase its rotational kinetic energy about the point of the contact. The angular speed $w$, and relative velocity of the rod with a length of $L$ can be defined as


\begin{equation}
w = \sqrt{\frac{3g}{L} \cos{\theta}}  \ \ \ \ \ \text{and} \ \ \ \ \  v_r = w L 
\end{equation}
where $g$ is the acceleration of the gravity and $\theta$ is the rotation angle \cite{DynamicBook}. Using the angular speed and relative velocity, it is possible to calculate the radar backscattering from a falling rod. Firstly, rod is modeled as a circular cylinder for two different lengths 0.75 and 2 meters as shown in Figure 3-(a) and (b). The normal backscattered RCS due to a linearly polarized incident wave from a circular cylinder of radius $r_c$ is  computed as
\begin{equation}
\sigma_c = \frac{2 \pi r_c L^2}{\lambda}
\end{equation}

Then, simulation can be completed by following the steps provided in Kinect micro-Doppler simulator. Resulting micro-Doppler signatures for two rods with different lengths are given in Figure 3-(c) and (d). As it can be easily noticed from the figures, longer rod has an increased speed compare to shorter rod. This information is valid for other non-rhythmic motions such as, sitting or bending.
 
Snapshots from animations derived from the same subject, but with different heights, are provided in Figures 4-(a) and (b). In Figure 4-(a), the subject's height is scaled down to 1.55 m, while in Figure 4-(b), the height is scaled up to 1.9 m. Corresponding micro-Doppler signatures are depicted in Figure 4-(e) and (f) for the same deviations. From micro-Doppler signatures, it may be seen that reflections from the right and left foot are more visually distinguishable for a taller subject due to the increased lower leg dimensions. Moreover, speed of the entire motion is relatively increased due to scaling along y-axis relative to the height change. This same relationship between height and micro-Doppler signature may be also observed in simulations based on the Boulic-Thalmann kinematic model for walking.
 
\begin{figure}[t]
	
	\begin{minipage}[b]{.48\linewidth}
		\centering
		\centerline{\includegraphics [width=1.7in] {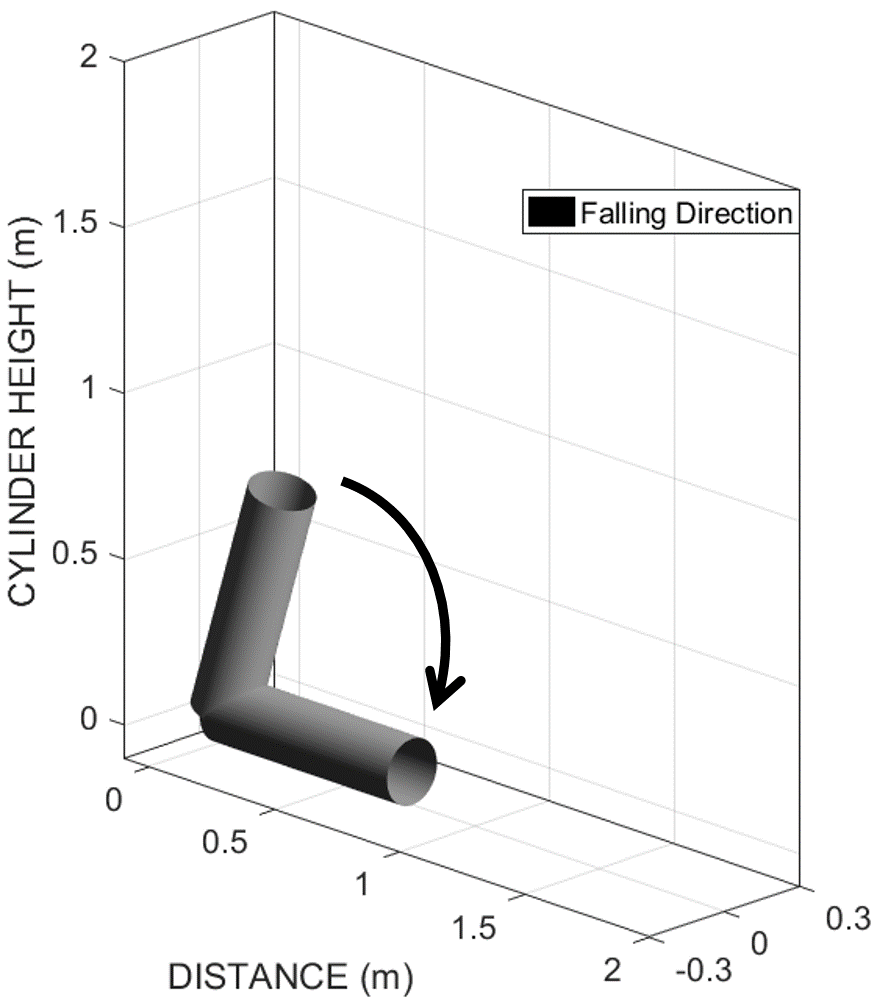}}
		\centerline{(a) }\medskip
	\end{minipage}
	\begin{minipage}[b]{.48\linewidth}
		\centering
		\centerline{\includegraphics [width=1.7in] {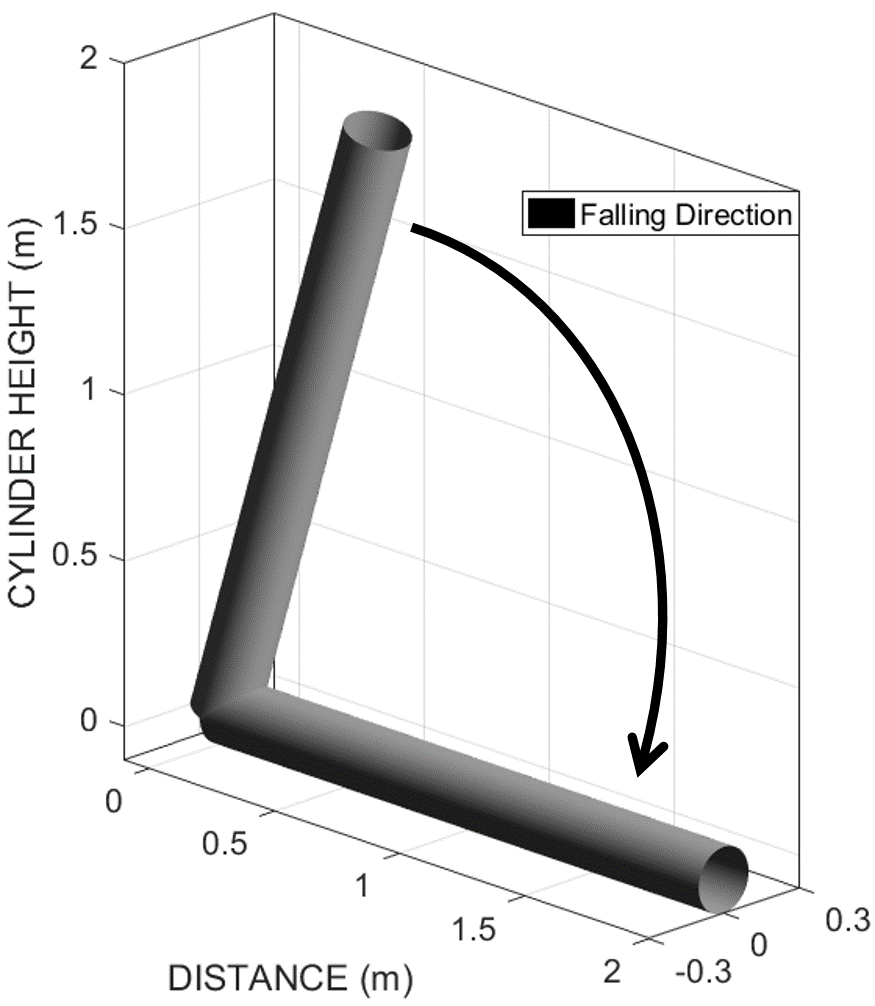}}
		\centerline{(b) }\medskip
	\end{minipage}
	\hfill
	\begin{minipage}[b]{0.48\linewidth}
		\centering
		\centerline{\includegraphics [width=1.78in, height=1.5in] {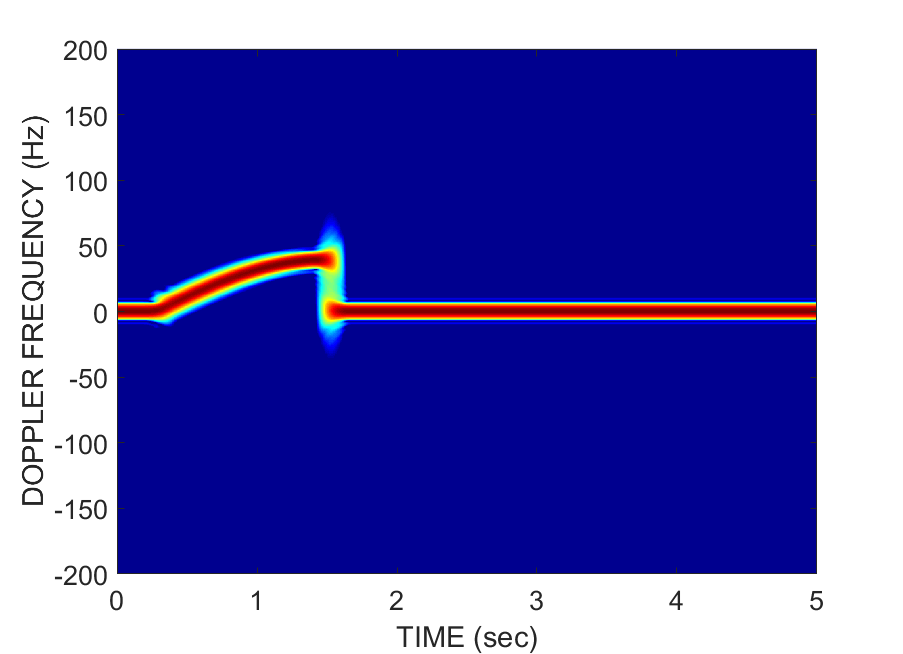}}
		\centerline{(c) }\medskip
	\end{minipage}
	\hfill
	\begin{minipage}[b]{0.48\linewidth}
		\centering
		\centerline{\includegraphics [width=1.78in, height=1.5in] {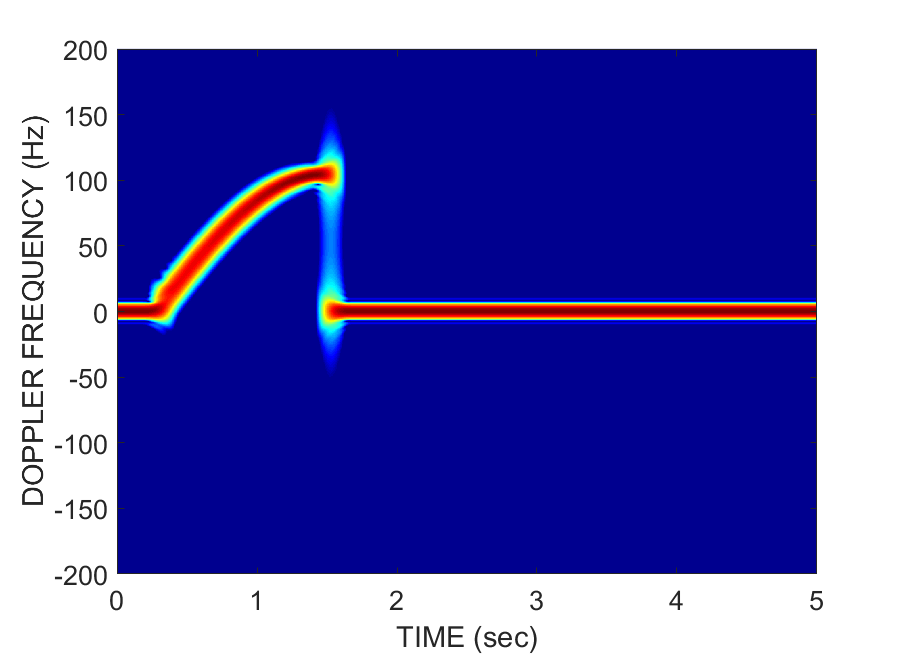}}
		\centerline{(d) }\medskip
	\end{minipage}
	
	\caption{Falling animation of (a) A short and (b) A long rod and corresponding falling micro-Doppler simulations for (c) A short and (d) A long length}
	\label{fig:res}
	
\end{figure}

\begin{figure*}[t!]
	\centering
	\begin{subfigure}[t]{.24\textwidth}
		\includegraphics [width=\columnwidth] {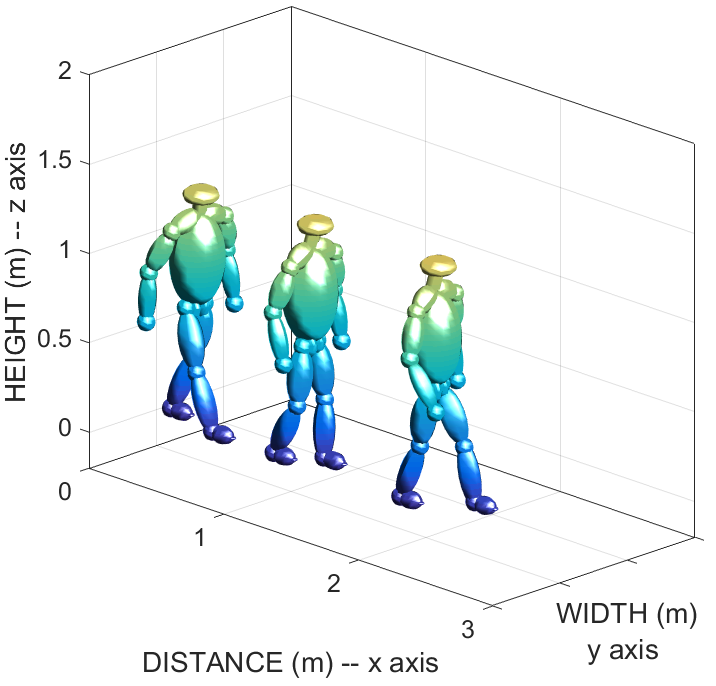}
		\caption{}
	\end{subfigure}
	\begin{subfigure}[t]{.24\textwidth}
		\includegraphics [width=\columnwidth] {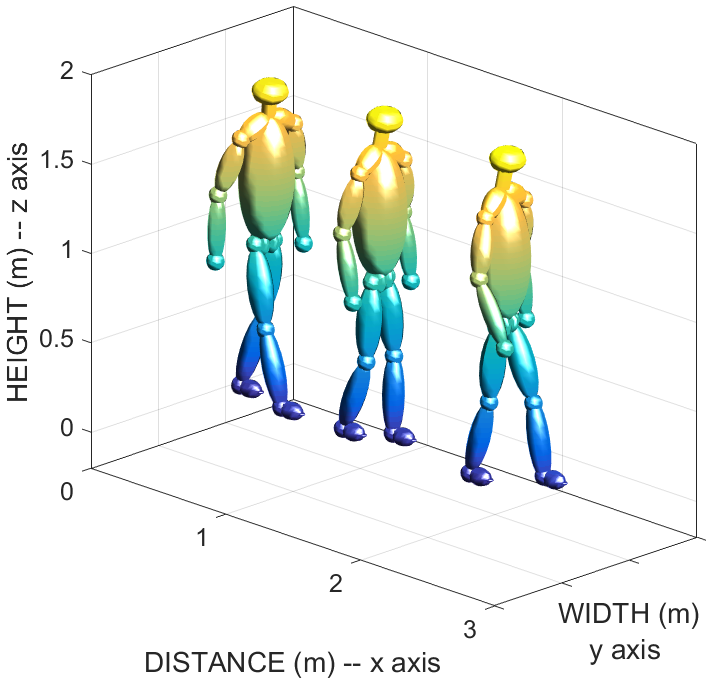}
		\caption{}
	\end{subfigure}
	\begin{subfigure}[t]{.24\textwidth}
		\includegraphics [width=\columnwidth] {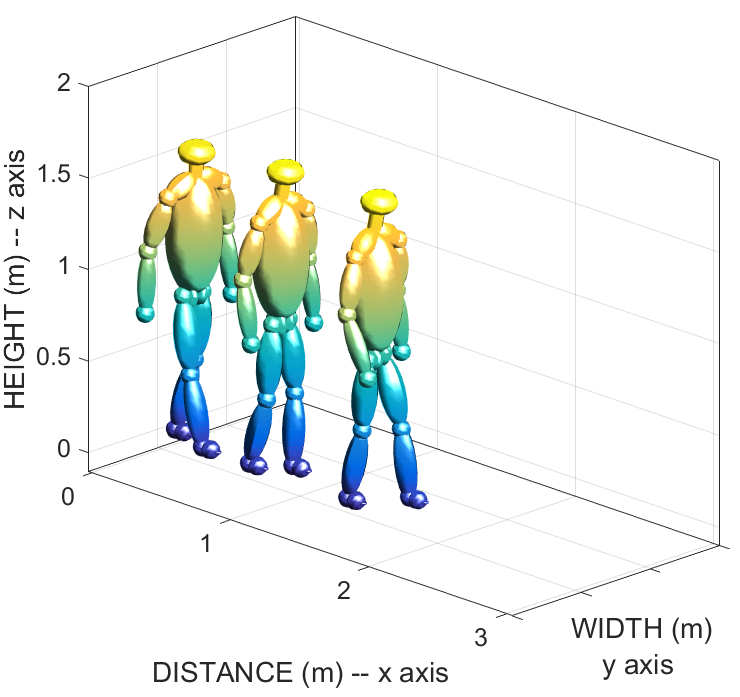}
		\caption{}
	\end{subfigure}
	\begin{subfigure}[t]{.24\textwidth}
		\includegraphics [width=\columnwidth] {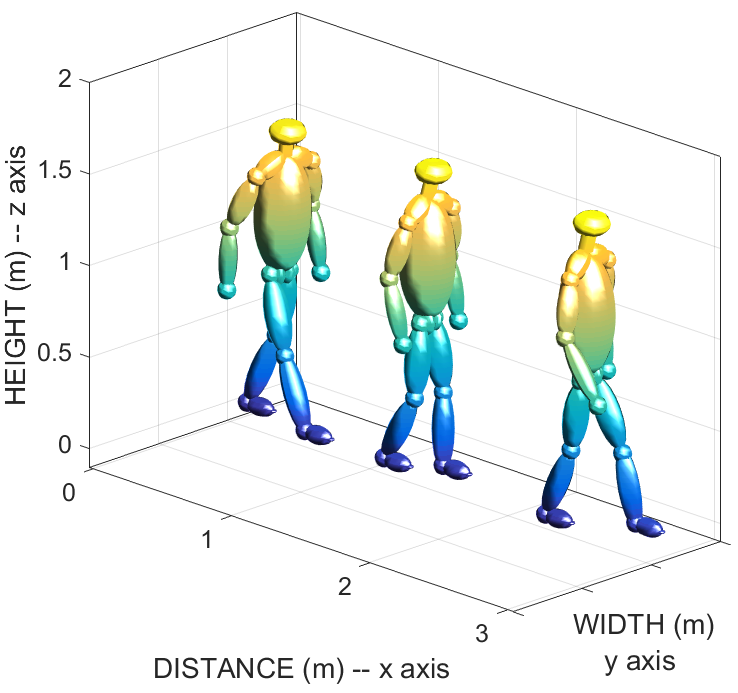}
		\caption{}
	\end{subfigure}
	\centering
	\begin{subfigure}[t]{.24\textwidth}
		\includegraphics [width=1.9in] {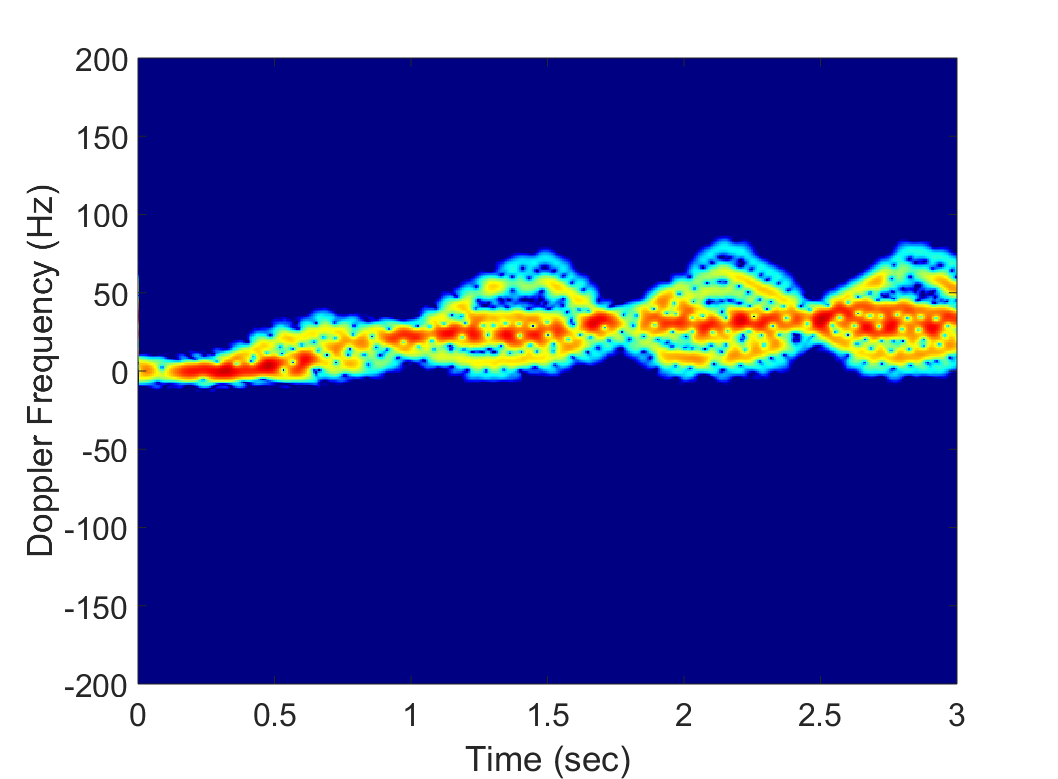}
		\caption{}
	\end{subfigure}
	\begin{subfigure}[t]{.24\textwidth}
		\includegraphics [width=1.9in] {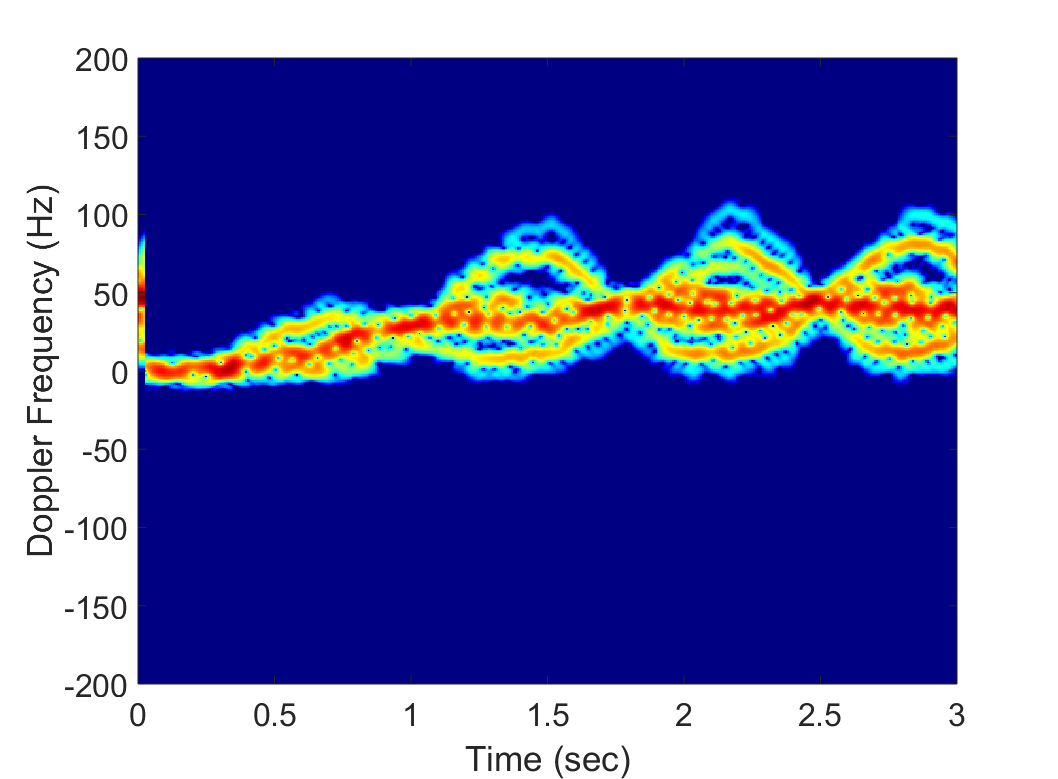}
		\caption{}
	\end{subfigure}
		\begin{subfigure}[t]{.24\textwidth}
			\includegraphics [width=1.9in,height=1.4in] {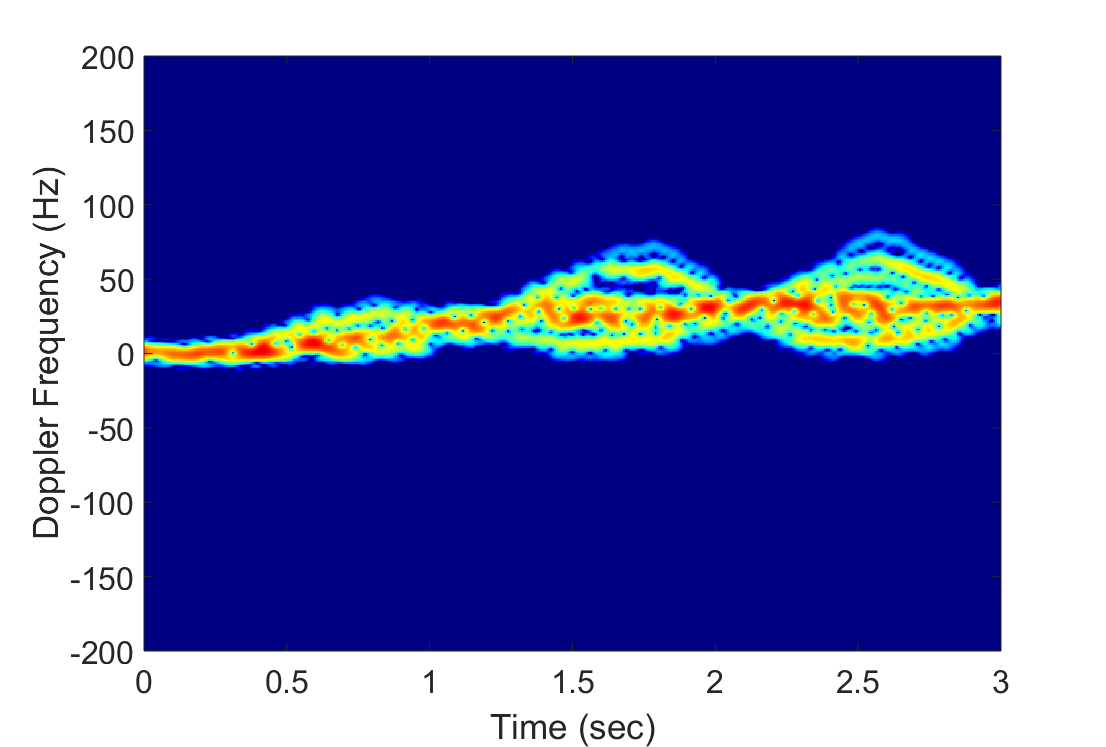}
			\caption{}
		\end{subfigure}
		\begin{subfigure}[t]{.24\textwidth}
			\includegraphics [width=1.9in,height=1.4in] {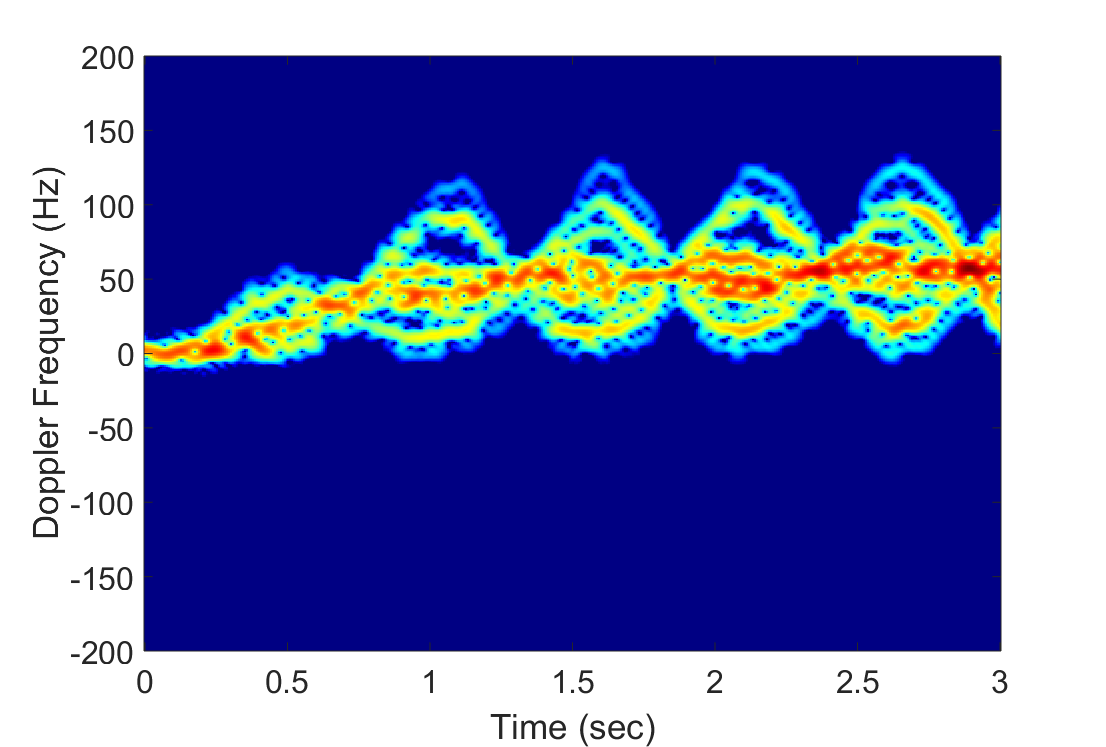}
			\caption{}
		\end{subfigure}
	\caption{Kinect-based animation results derived from one  data for (a) A short subject (b) A tall subject (c) Subject with a slow stride rate, (d) Subject with a fast stride rate and resulting micro-Doppler signatures for 15 GHz CW radar (e) Short (f) Tall (g) Slow, and (h) Fast}\label{label-b}
\end{figure*}

A second parameter that greatly influences the micro-Doppler signature is the speed of the subject.  In the Boulic-Thalmann model, the speed of motion can be varied by simply changing one parameter, cycle length or duration ($d_c$). These fundamental spatial and temporal characteristics of a walk are defined as

\begin{equation}
l_c = 1.346 \sqrt{v_r}  \ \ \ \ \ \text{and} \ \ \ \ \  d_c = \frac{l_c}{v_r} 
\end{equation}

The first spatial characteristic formula is obtained from the normalization formula defined in \cite{WalkingBook}. Note that, normalization constant is determined specifically for human gait by examining lots of experimental studies. The rest of the parameters change accordingly through kinematic relationships between different body parts. However, completing this task in Kinect-based MOCAP data can be challenging due to fact that the acquired position data already contains a speed factor. Therefore, another operation along x-axis (distance) is required. We start by manipulating the sample frequency of the raw Kinect data which changes the stride rate and speed of the motion. In Figure 4-(c) and (d), animations derived from one sample are provided for two different speeds. It is evident that a faster subject travels longer distance than a slower subject within the same time interval. The micro-Doppler signatures are also depicted in Figure 4-(g) and (h) for the same derivations. Note that faster subjects exhibit shortened cycles than slower subjects within the same time interval.

Another important consideration in our simulation methodology is about how to discard extreme cases that lead to signature overlapping and thus cause confusion, such as those associated with slow running and fast walking. Variants of these two motions might possibly reside in the same frequency bands as a consequence of changing the speed of motion. Therefore, to prevent such unwarranted overlapping, we determined the highest and slowest speeds as well as the corresponding Doppler frequencies (15 GHz center frequency) for 7 different human activities, and then limited the diversified signatures accordingly.

\begin{figure}[t]
	\includegraphics[width=3in]
    {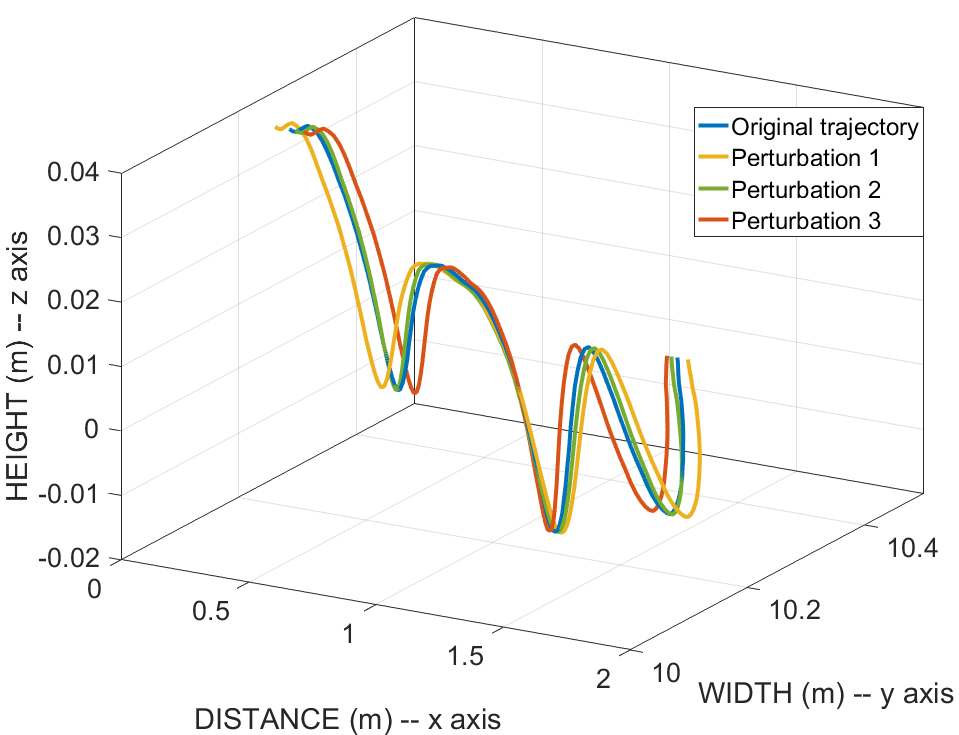}
	\caption{Original trajectory of the left arm and perturbed trajectories}
	\label{fig:validacc30db4layers}
\end{figure}

\subsubsection{Parameterization of Individual Joints}
The last step of the diversification methodology complements scaling and focuses on the individual joint data, such as left and right leg, right and left arm, and head. The main idea is to parameterize the Kinect raw distance data of the different joints separately. Then, by just perturbing the coefficients of the constructed models, it is possible to create class variations. Limited alterations of the model parameters  manifest itself in the style of how the motion is performed.  

Parameterization of the joint can be done in several ways, most easily using curve fitting models. To this end, we employed different types of curve fitting models, such as sinusoidal, Fourier series (harmonic), polynomial, linear interpolation, and so forth. Given the periodic nature of the  joint data and by examining the goodness of the constructed models, the Fourier series was determined to be the best suited parameterization model for the underlying problem. This model also provides a good fit for non-period range trajectories, which are mostly encountered in non-rhythmic motions, such as falling and sitting. The Fourier series model describes the given Kinect range data, $x$ defined between $1$ to $m$, as a sum of sine and cosine functions. Resulting model can be represented in the trigonometric form as
\begin{equation}
f(x,\mathrm{a}_j,\mathrm{b}_j)=\mathrm{a}_{0}+\sum_{j=1}^{n}\mathrm{a}_{j} \cos(jwx) + \mathrm{b}_{j} \sin (jwx), 
\end{equation}
where $\mathrm{a}_0$ models a constant term in the data and is associated with the cosine term for $j=0$, $w$ is the fundamental frequency of the signal, and $n$ ($0<n<9$) is the number of terms (harmonics). Non-linear least squares method is employed using Levenberg-Marquardt algorithm \cite{levenberg_method_1944} to find the coefficients that gives the best fitting curve and objective function is defined as
\begin{equation}
\{\mathrm{\tilde{a}_{j}},\tilde{b}_{j}\} = {\text{argmin}} \ \ \sum_{k=1}^{m} \norm{x_{k}-f(x_{k},a_j,b_j)}^2  \ (j=1, 2, ..., n)
\end{equation}

The Fourier series model in (10), provides $2n$ coefficients, which we can alter depending on the number of harmonics used. To prevent demolishing the joint trajectory entirely, and preserve the underlying information of the joint data, we only changed the  n-dual harmonic coefficients $[(a_1, b_1), ..., (a_n, b_n)]$, one pair at a time, and within a 10\% range value. 

Note that, depending on the motion that we put into the parameterization model, the number of Fourier coefficients that algorithm decides upon would change. Initially, we would like to use as many coefficients as possible in our model because that would provide a better fit to the variations. Parameterization starts with 8 Fourier coefficients. Then, depending on the goodness of the fit and the value of the coefficients, the algorithm automatically increases or decreases the number of coefficients. Afterwards, it again computes the goodness of the fit and checks the values. This process is repeated until a pre-determined fit goodness is met. 

Figure 5 depicts the original trajectory of the left arm and the perturbed trajectories for walking. It may be observed that perturbation does not completely demolish the underlying structure of the trajectory information. Some constraints are also imposed to make the methodology more consistent with the kinematics of different motions. For example, when a harmonics's dual coefficient pair is changed for different joints, the algorithm automatically compares the altered arm and leg lengths with the original lengths to determine whether alteration is kinematically possible. We again emphasize that Fourier-series-based modeling and parameter adjustments do not provide exact representations of possible kinematic motions, but rather approximate them with  different degrees.

\subsection{Validation of the Proposed Methodology}

Validation of the diversification algorithm is accomplished by visually examining the output animations of the extreme examples and computing the structural similarity index (SSI) for intra- and 2-D correlation coefficients for inter-class samples \cite{SSIMref}. For visual observations, we determined the maximum and minimum speeds and height values of the subjects so as to remove any extreme samples from the diversified database. The  animations and micro-Doppler signatures were then examined to ascertain the physical feasibility of the motion, specifically at high scaling and perturbation values. Since it is not possible to conduct the visual observations for numerous variants of each motion, the SSI and 2D correlation coefficients are computed to measure the similarity between inter- and intra-class samples, respectively. Whereas the former represents similarities among members of the same class, the latter pertains to samples of different classes. 

First, the 2D correlation coefficients can be computed to obtain the inter-class similarity map of the ground truth samples as depicted in Figure 6-(a). Note that, each class has their unique cluster in this map. However, falling and sitting samples seem to receive similar correlation values, which is anticipated due to the similar kinematic structure of these two motions. Other than falling and sitting, only a few samples in the overall map receive higher similarity indices for samples from different classes. 

The SSI can be viewed as a quality metric for images being compared, provided the other image is regarded as of original quality. This index is based on the computation of three terms, namely the luminance term, the contrast term and the structural term. The overall index is a multiplicative combination of the three terms:
\begin{equation}
SSI(x,y) = \frac{2(\mu_x \mu_y +C_1) (2\sigma_{xy}+C_2)}{(\mu_x^2+\mu_y^2+C_1)(\sigma_x^2+\sigma_y^2+C_2)}
\end{equation}
where $\mu_x$, $\mu_y$, $\sigma_x$, $\sigma_y$, and $\sigma_{xy}$ are the local means, standard deviations, and cross-covariance for images x, y. Also, $C_1$ and $C_2$ are defined as the regularization constants for the luminance and contrast, respectively. We examined the similarities between the original micro-Doppler signatures and the variants generated through our diversification algorithm. To generate variance from the limited number of original samples, the similarity between the original and variant images must assume a value less than 1. Too low of a value, however, can indicate a large divergence and have an adverse effect on classification accuracy. SSI is computed for the original images and generated variants as depicted in Figure 6-(b) for three classes (walking, running, and limping). For each iteration involving height, each class exhibits patterns in SSI that are unique to its class, indicating that the variants generated are consistent with overall class characteristics.


\begin{figure*}[t]
	\centering
	\begin{subfigure}[t]{.45\textwidth}
		\includegraphics [width=3in,height=2.0in] {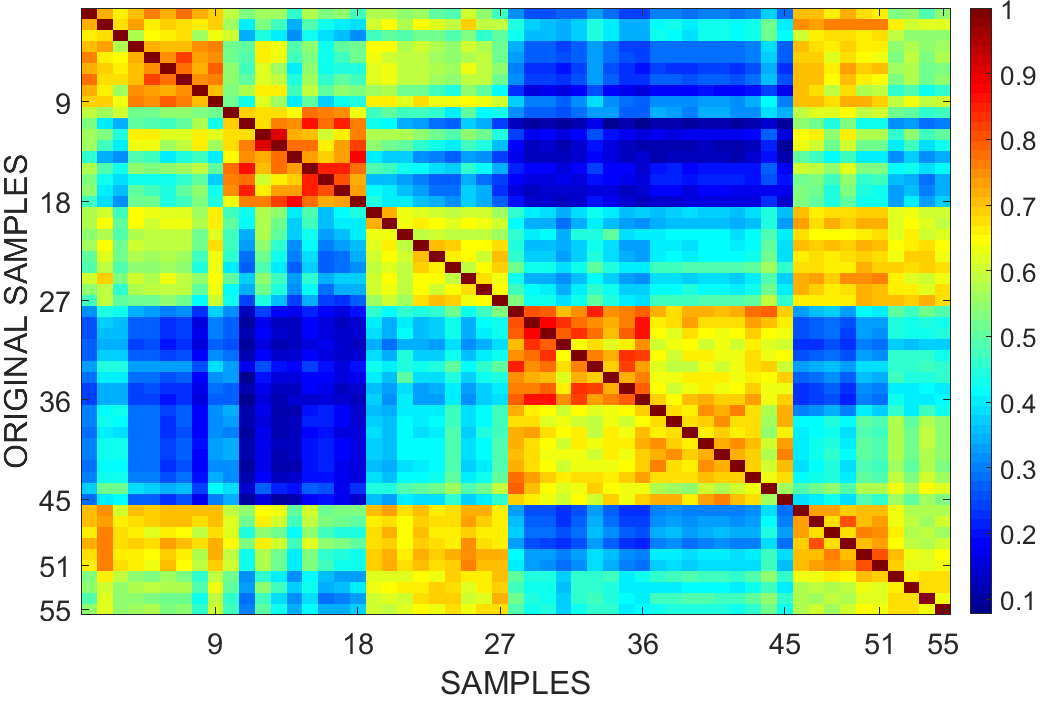}
		\caption{}
	\end{subfigure}
	\begin{subfigure}[t]{.45\textwidth}
		\includegraphics [width=3in,height=2.0in] {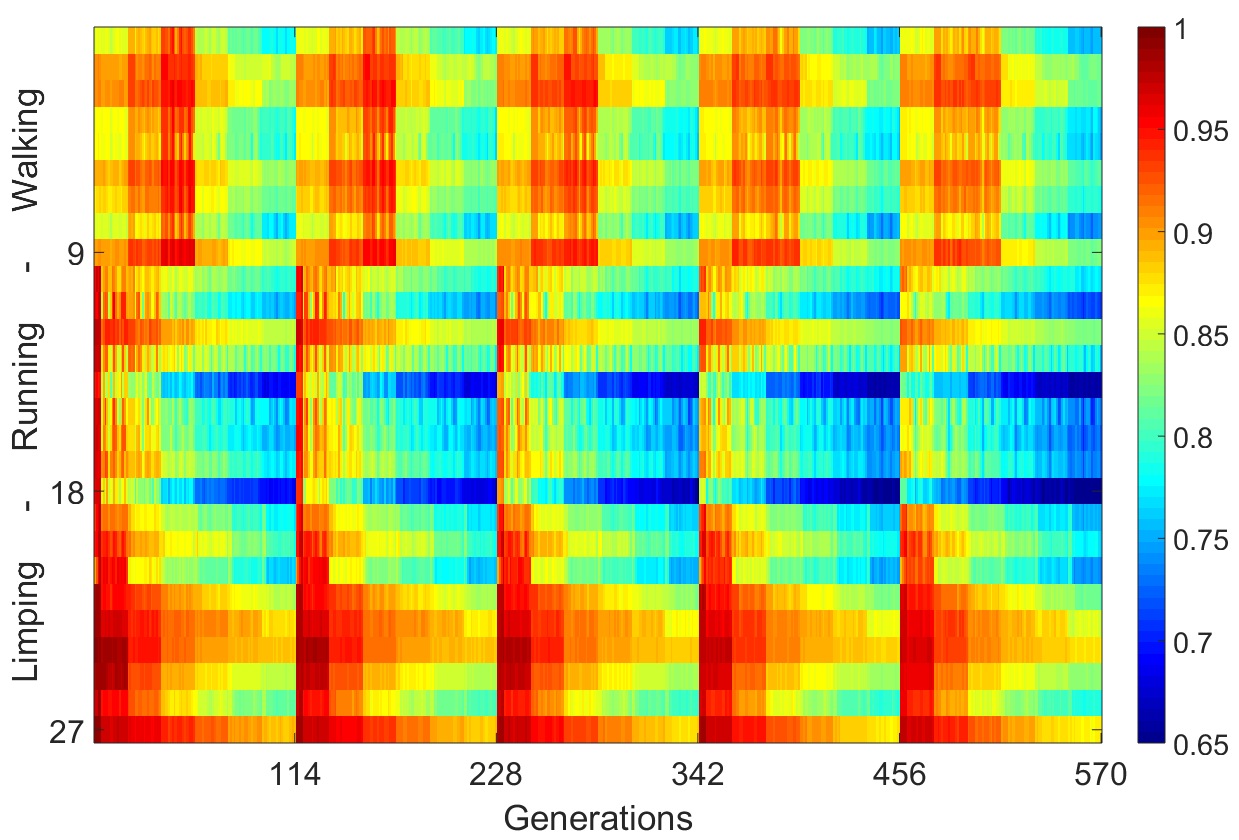}
		\caption{}
	\end{subfigure}

	\caption{Validation of the proposed signature diversification methodology: (a) Inter-Class comparison between ground-truth examples (1-9 Walking, 9-18 Running, 18-27 Limping, 27-36 Falling, 36-45 Sitting, 45-51 Cane, and 51-55 Walker) and (b) Intra-class similarity comparison (x-axes depicts the height iterations: 1-114 first, 114-228 second, 228-342 third, 342-456 fourth, and 456-570 fifth)}\label{label-b}
\end{figure*}



\section{Experimental Test Dataset}
The test set used in this work comprises entirely real radar measurements conducted in an indoor laboratory at TOBB University.  An NI-USRP 2922 model software-defined radio, mounted 1 meter above the ground, was used to transmit a 4 GHz CW signal. Two SAS-571 horn antennas were placed on either side of the USRP to yield an approximately monostatic configuration.  Test subjects conducted seven different human activities at a range from the radar varying between 1 meter and 5 meters.  All experiments were conducted in alignment with the radar line-of-sight.   

Two datasets were formed for testing the proposed transfer learning approach:  a 7-class data set, and an 11-class dataset.  For the 7-class dataset, the activities enacted along with total number of measurements per class are as follows: walking (71), jogging (72), limping (104), walking while using a cane (123), walking while using a walker (121), falling forward from a standing position (53), and sitting (50).  This data set was augmented with data from four additional classes to form the 11-class data set; namely, creeping (56), crawling (74), using a wheelchair (149), and walking with crutches while one leg is bent at the knee (74).  

Micro-Doppler signatures are represented as spectrograms, computed using a hamming window with length of
2048 samples, 4096 FFT points, and 128 samples overlap. Each spectrogram was then cropped to a duration of
4 seconds, converted to grayscale and saved as an image. To
reduce dimensionality, the resulting images were then downsampled
from a size of 656x875 pixels to 90x120 pixels or 327x436. 

\section{CNNs Trained on Measured Radar Data}

\begin{table*}[!t]
	\centering
	\caption{Validation accuracy (VA) and test accuracy (TA) as a function of DNN parameters for 7-class case for a CNN trained on measured data and on diversified data (overfitting cases highlighted)}
    \label{t2}
	\begin{tabular}{|c|c|c|c|c|c|c||c|c|c|}
		\hline
		 & \multicolumn{6}{c||}{\textbf{Trained with Measured Data Only}} 
      & \multicolumn{3}{c|}
         {\textbf{Trained with Diversified Data}}
         \\
      
    & \multicolumn{3}{|c|}{\textbf{90 x 120 Pixel Input}} & \multicolumn{3}{|c||}{\textbf{327 x 436 Pixel Input}} & \multicolumn{3}{|c|}{\textbf{90 x 120 Pixel Input}} \\
\hline
\textbf{No. of Conv. Layers}&  \textbf{VA Max@Epoch} & \textbf{VA @500} & \textbf{TA} & \textbf{VA Max@Epoch} & \textbf{VA @500} & \textbf{TA}  & \textbf{Train Acc.} & \textbf{VA} &  \textbf{TA}\\
 \hline\hline
\textbf{2} 
   & 86\% @ 30 & 83\% & 84\% & 85\% @ 30 & 79\% & 80\%  & 99\% & 95\% & 93\% \\
 \hline
 
\textbf{4}
  & 86\% @ 65 & 86\% &87\% & 85\% @ 25 & 84\% &  83\%  & 100\% & 94\% & 94\% \\
 \hline
 
 \textbf{7}
   &  81\% @ 151 &  81\% & 77\%& 85\% @ 40 & 84\% & 79\%  & 99\% & 96\% & 96\% \\
 \hline
 
\textbf{11}
  & \cellcolor{yellow!70} 83\% @ 132 & \cellcolor{yellow!70} 79\% & \cellcolor{yellow!70} 80\% & 82\% @ 60 & 82\% & 81\%  & 95\% & 90\% & 91\% \\
 \hline

\textbf{15}
  & \cellcolor{yellow!70} 80\% @ 116  & \cellcolor{yellow!70} 76\% & \cellcolor{yellow!70} 74\% & \cellcolor{yellow!70} 72\% @13  & \cellcolor{yellow!70} 64\% & \cellcolor{yellow!70} 70\%  & 88\% & 84\% & 85\% \\
 \hline

\textbf{20}
  & \cellcolor{yellow!70} 74\% @ 97  & \cellcolor{yellow!70} 68\% & \cellcolor{yellow!70} 67\% & \cellcolor{yellow!70} 65\% @ 30  & \cellcolor{yellow!70} 61\% & \cellcolor{yellow!70} 61\%  & \cellcolor{yellow!70} 100\% & \cellcolor{yellow!70} 78\% & \cellcolor{yellow!70} 80\% \\
 \hline

\textbf{25}
  & \cellcolor{yellow!70} 64\% @74   & \cellcolor{yellow!70} 52\% &  \cellcolor{yellow!70} 49\% & \cellcolor{yellow!70} 69\% @21   & \cellcolor{yellow!70} 43\% & \cellcolor{yellow!70} 42\%   & \cellcolor{yellow!70} 100\% & \cellcolor{yellow!70} 63\%  &\cellcolor{yellow!70} 64\%  \\
 \hline

\textbf{30}
  & \cellcolor{yellow!70} 14\%   &  \cellcolor{yellow!70} 14\% &  \cellcolor{yellow!70}14\% & \cellcolor{yellow!70} 14\%    & \cellcolor{yellow!70} 14\% & \cellcolor{yellow!70} 14\%  & \cellcolor{yellow!70} 100\% & \cellcolor{yellow!70} 14\%  & \cellcolor{yellow!70}14\% \\
 \hline

\end{tabular}
\end{table*}

CNNs \cite{lecun1990handwritten} have recently gained in popularity as the state-of-the-art in image classification, blossoming in the fields of computer vision and natural language processing, where millions of data samples are easily obtained. Within the CNN architecture, there are perhaps thousands of hyperparameters that may be adjusted to find the right combination leading to the best representation of the data; e.g., dimensionality of input (size of image), the number of layers in the network (depth), number of neurons per layer, number of filters per layer (width), spatial size of filters, type of activation function, and size of pooling operation. 

When pooling layers are utilized, the greater the input dimensionality, the deeper a CNN may be constructed.  Increasing dimensionality, however, does not necessarily imply greater performance, while it does increase the overall computation time for the network. Filter size and output feature map size quadratically increase the time complexity \cite{He15}, while the depth and the input dimension results in a linear increase. Increasing the depth, however, has a much greater benefit to performance than other DNN parameters.  In feedforward neural networks, it has been theoretically shown \cite{eldan16} that increasing depth by just 1 is exponentially more valuable than increasing the number of neurons per layer, and that deeper networks can also have lower errors than wider but shallower counterparts \cite{telgarsky}.  In CNNs, depth has been shown to be more significant than width or filter size \cite{He15}. 

Thus, in this work, we focused on the impact of depth and input dimensionality on performance using the 7-class measured dataset described in Section III.  The measured data is separated into subsets of 80\% for training and 20\% for testing using 5-fold cross-validation. The validation test set is derived from random selection of 20\% of the training dataset. The classification results attained are shown in Table \ref{t2}, and were generated using a randomly initialized CNN with 32 filters on each layer with a size of 3x3, a rectified linear unit (ReLU) activation function and two fully connected layers with 150 neurons in each layer.  A 2x2 max pooling was used for all CNN depths except that with 11-layers. After the first fully connected layer, a dropout operation is applied with a probability of 50\% to mitigate overfitting. Note that in Table II, the maximum validation accuracies are typically observed much sooner than the final 500th epoch, while a tell-tale sign of overfitting is divergence between training and validation accuracies.  Thus, in many cases, overfitting can be thus prevented by early stopping the training process.

Cases that exhibited problems of overfitting are highlighted in Table II.  While overfitting was observed at greater network depths for the larger input size 437x436, we found that the validation and test accuracies using the larger input size were low than that of the smaller image, given the same network depth.  Thus, an input dimension of 90x120 was selected.

Through comparison of validation and test accuracies, the generalization performance of the DNN may also be assessed.  Significant differences between the validation and test accuracies, such as incurred in the 7 convolutional-layer CNN, are an indication that the network cannot generalize to new test cases; i.e., individual gate or body type.  For the 7-class problem, a 4-convolutional layer CNN, taking inputs of 90x120 pixels yields the highest test accuracy with no generalization loss.

\section{DNNs Trained on Diversified Data} 

\subsection{CNNs}
The limitations in classification accuracy and network depth may be overcome by using a larger dataset spanning the in-class variations expected from different target profiles.  In this work, such a diversified dataset of 32,000 samples is generated using the methodology outlined in Section II from 55 original Kinect-based MOCAP measurements of 5 test subjects collected in the Radar Imaging Laboratory at Villanova University.  The diversified signatures are used for initial pre-training of CNN weights, while subsequently 474 measured micro-Doppler signatures are used for fine-tuning.  The remaining 120 measured signatures are used for the test dataset.  Note that the test subjects from Villanova University used in acquiring MOCAP measurements are entirely different from those used in collection of real measurements at TOBB University.  The training, validation, and test accuracy achieved for CNNs with different depths and an input image size of 90x120 pixels are given in Table II.

Comparing the results for measured and diversified training data given in Table II, it may be observed that use of the proposed diversified dataset in the training process has in general resulted in higher classification accuracies - a 9\% improvement when comparing the best performing architectures.  Moreover, overfitting does not become significantly large until greater depths, e.g. 20 convolutional layers.  However, it is important to notice that at 11 and 15 layers - before any overfitting has been observed - a drop in training accuracy, which parallels a similar drop in validation accuracy, occurs.  Degradation in training accuracy (increased training error) is a problem that has been observed as network depth increases:  first classification accuracy is saturated, followed by a rapid decrease. To remedy this problem, utilization of residual units have been recently proposed \cite{He16_ResNet} as a means to improve optimization, and, thus, accuracy as a function of depth. 

\subsection{Residual Learning}
Deep residual networks are comprised of building blocks, which rather than computing the original mapping of $y_l(x) := h(x_l)$, compute the residual mapping of 
\begin{equation}
y_l = h(x_l) + F(x_l,W_l),
\end{equation}
\begin{equation}
x_{l+1} = f(y_l),
\end{equation}
where $x_l$ is the input to the $l^{th}$ residual unit (RU), $W_{l}={W_{l,k|1 \leq k \leq K}}$ is a set of weights and biases for the $l^{th}$ RU, K is the number of layers in a RU; $F$ is a residual function, e.g. a stack of two convolutional layers; $h(x_l) = x_l $ is an identity mapping that is performed by using a shortcut path, and $f$ is an activation function (e.g., ReLU). In addition, batch normalization (BN) layers are used, where batches are standardized (zero mean/unit variance) throughout the network to reduce the interval co-variance shift, which allows larger learning rates to be used \cite{ioffe2015batch}. Shortcut connections can thus be created within the network by simply implementing an \textit{identity} mapping.  By driving subsequent layers with the residual (i.e., the input $x$ as well as output of previous layer), the network is forced to effectively learn something new beyond that already embodied by previous layers. Thus, another advantage of residual learning is that very deep networks can be constructed without worrying about whether the network is "too deep" - if adding layers gives no benefit, residual blocks can learn the identity mapping and thus do no harm to performance.

\begin{figure*}[t]
	\includegraphics[width=16cm]{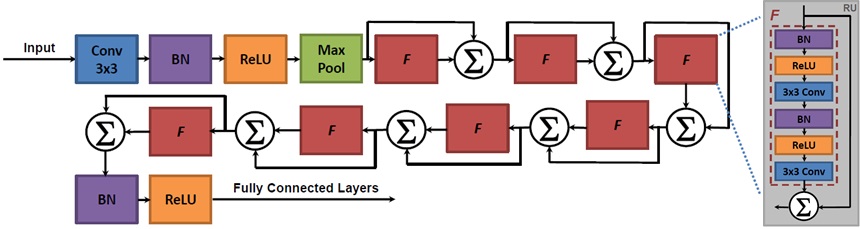}
	\caption{Proposed 15-convolutional layer residual learning DivNet architecture.}
	\label{fig:DivNet}
\end{figure*}

In this work, we utilize a modified RU \cite{he2016identity}, which is more easily trainable and possesses improved generalization properties.  In the modified RU, the activation function is now placed before the final addition, as opposed to afterwards.  The resulting overall residual transfer learning network (called DivNet), which we propose to be trained on  diversified micro-Doppler signatures, includes 7 residual units and is comprised of a total of 32 layers (with 15 convolutional) layers, as shown in Figure \ref{fig:DivNet}.


\begin{table}[t]
	\centering
	\caption{Classification performance for 7-class case of residual DivNet pre-trained on 90x120 pixel diversified data.}
	\begin{tabular}{|c|c|c|c|}
\hline
\textbf{No. of Conv. Layers}&  \textbf{Train Acc.} & \textbf{VA} & \textbf{TA}  \\
 \hline\hline
\textbf{5} 
   & 100\%  & 96\% & 96\%   \\
 \hline
 
 \textbf{10} 
   & 100\%  & 95.6\% & 96\%   \\
 \hline
 
 \textbf{15} 
   & 100\%  & 98\% & 97\%   \\
 \hline
 
 \textbf{20} 
   & 100\%  & 95\% & 94\%   \\
 \hline
 
 \textbf{25} 
   & \cellcolor{yellow!70} 100\%  & \cellcolor{yellow!70} 78\% & \cellcolor{yellow!70} 73\%   \\
 \hline
 
 \textbf{30} 
   & \cellcolor{yellow!70} 100\%  & \cellcolor{yellow!70} 14\% & \cellcolor{yellow!70} 14\%   \\
 \hline
\end{tabular}
\label{t3}
\end{table}

The improvement in training and classification accuracy attainable using residual transfer learning on the diversified dataset is shown in Table \ref{t3}.  With the utilization of residual units, the problem in training accuracy degradation has been resolved, and overfitting is not observed until a depth of 25 convolutional layers.  Validation accuracy has also been further improved by 1\% to 97\% for 15 convolutional layer DivNet (a.k.a. DivNet-15, in short).  Note that the performance gains achieved by using the diversified data (an improvement of 9\% for CNNs, as shown in Table II) significantly surpasses the accuracy improvement caused by residual learning for this particular dataset; in combination, however, DivNet has the ability to offer the construction and training of arbitrarily deep neural networks for micro-Doppler classification.

\section{Advantages of DivNet}

In this section, a closer look at the benefits offered by initial pre-training on diversified, simulated data will be made.  In particular, the proposed approach will be compared to using the current conventional approach of training with a small set of measured data, and a recently proposed approach of using transfer learning with initial pre-training on optical imagery, such as the ImageNet database.

Transfer learning is a technique that has been proposed for domain adaptation when a little or no labeled data is available relating to the targeted classes. The network is first initialized using labeled data from a different source domain; then, the small set of task-related labeled data is used to fine tune initialized parameters prior to classification of the test set.  In radar micro-Doppler literature, the 16-layer VGGnet \cite{vggnet} has been pre-trained on the ImageNet \cite{imagenet} database, comprised of 1.5 million RGB images, while fine-tuned with just 625 measurements to yield a classification accuracy of 80.3\% in distinguishing 5 different types of swimming \cite{park16}.  However, DNNs that perform well on optical imagery, do not necessarily have the comparable level of performance on radar data.  For example, on ImageNet, the 22-layer CNN named GoogleNet \cite{googlenet} yielded surperior performance than VGGnet - yet in a recent study \cite{seyfiogluGSRL2017} comparing performance on micro-Doppler data, VGGnet was found to consistenly outperform GoogleNet.  Thus, in this section we compare the performance of DivNet-15 with transfer learning from optical imagery using VGGnet and a 50-layer version of the deepest, most advanced network constructed to date, Microsoft's  residual neural network named ResNet \cite{resnet}, and a 4-layer CNN trained on 474 samples of measured data (described in Section IV). 

\subsection{Bottleneck Feature Performance Comparison}
To evaluate the differences in network initialization offered by the different networks (VGGnet, ResNet-50 and proposed DivNet-15) and transfer domains (optical versus simulated, diversified RF), the classification performance of bottleneck features alone is examined. Bottleneck features refer to the last activation maps before the full-connected layers. In order to examine the performance of bottleneck features, the previously trained fully-connected layers are removed from the network and, instead, two randomly-initialized fully-connected layers, each containing 150 neurons are added, followed by a softmax classifier with 7 neurons. Each layer up to the fully-connected layers is frozen; thus, weights of the convolutional blocks do not change by back-propagation. With fixed activation maps, the randomly-initialized fully-connected layers are trained with measured micro-Doppler data by using stochastic gradient descent with a learning rate of 0.001 and batch size of 50. Using this procedure allows us to isolate the impact of initial pre-training of network weights on performance.

\begin{figure*}[t!]
\centering
	\begin{subfigure}[t]{.32\textwidth}
    \centering
		\includegraphics [width=2.4in] {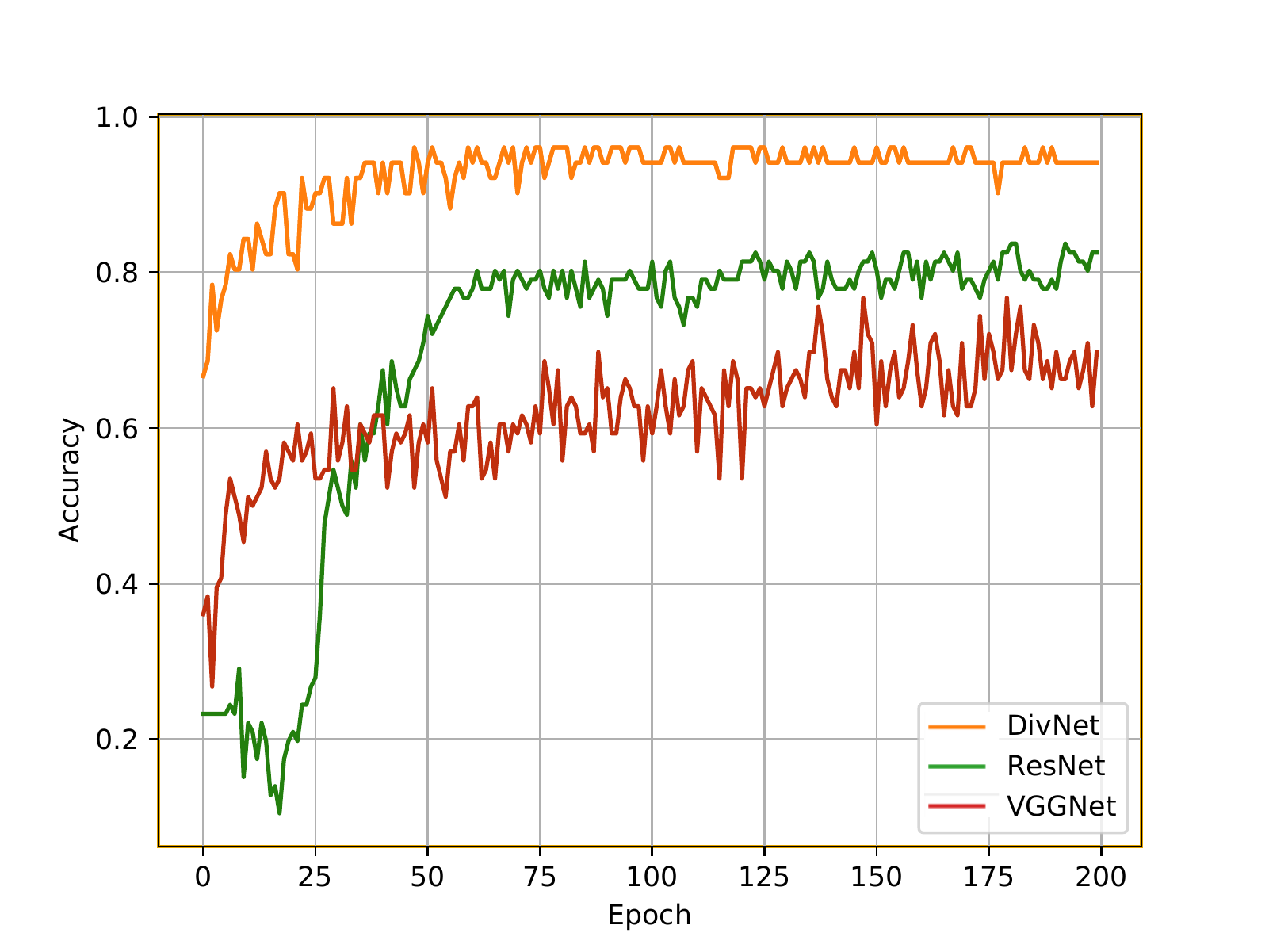}
        \caption{}
	\end{subfigure}
	\begin{subfigure}[t]{.32\textwidth}
    \centering
		\includegraphics [width=2.4in] {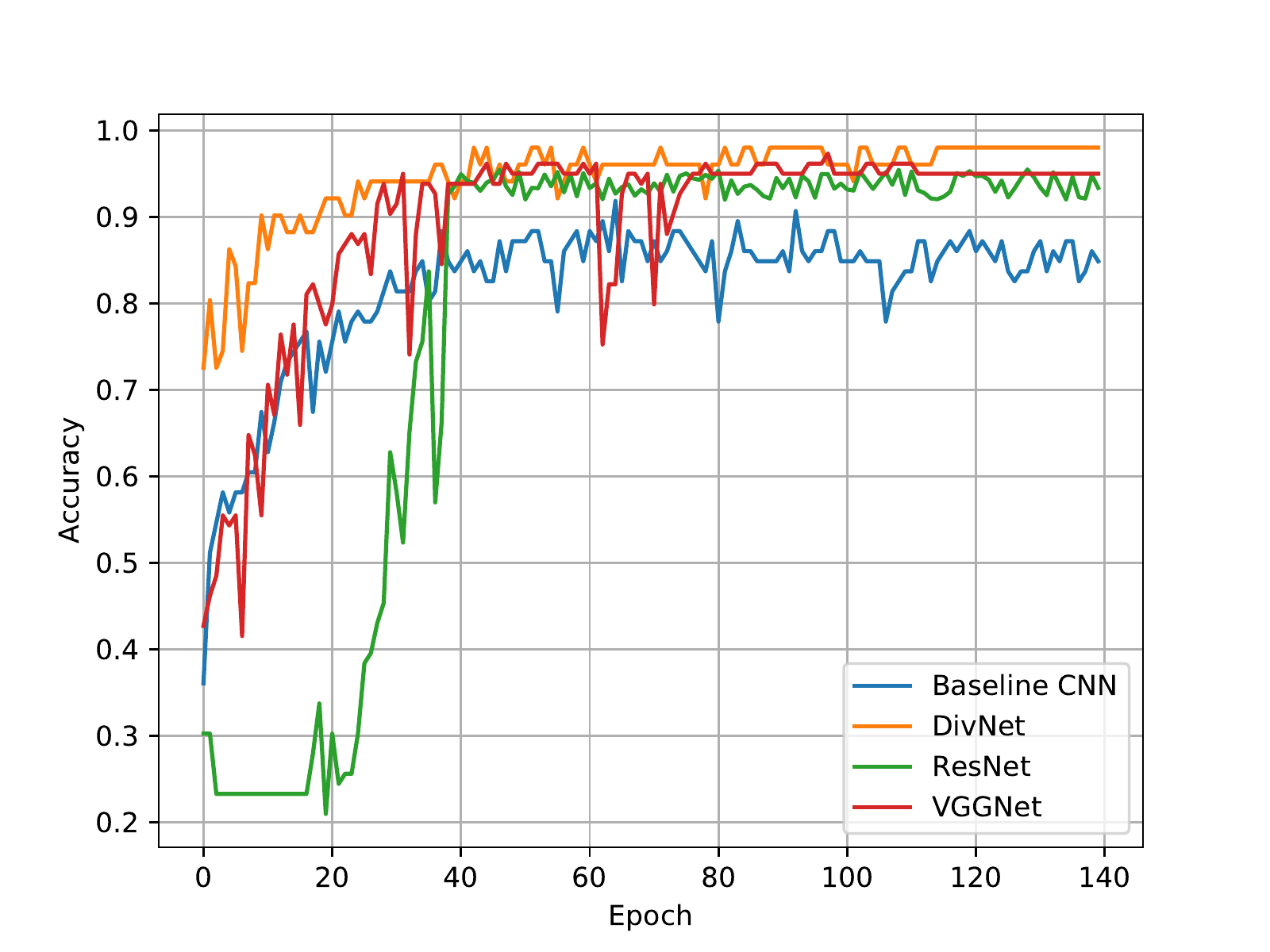}
        \caption{}
	\end{subfigure}
    \begin{subfigure}[t]{.32\textwidth}
    \centering
		\includegraphics [width=2.4in] {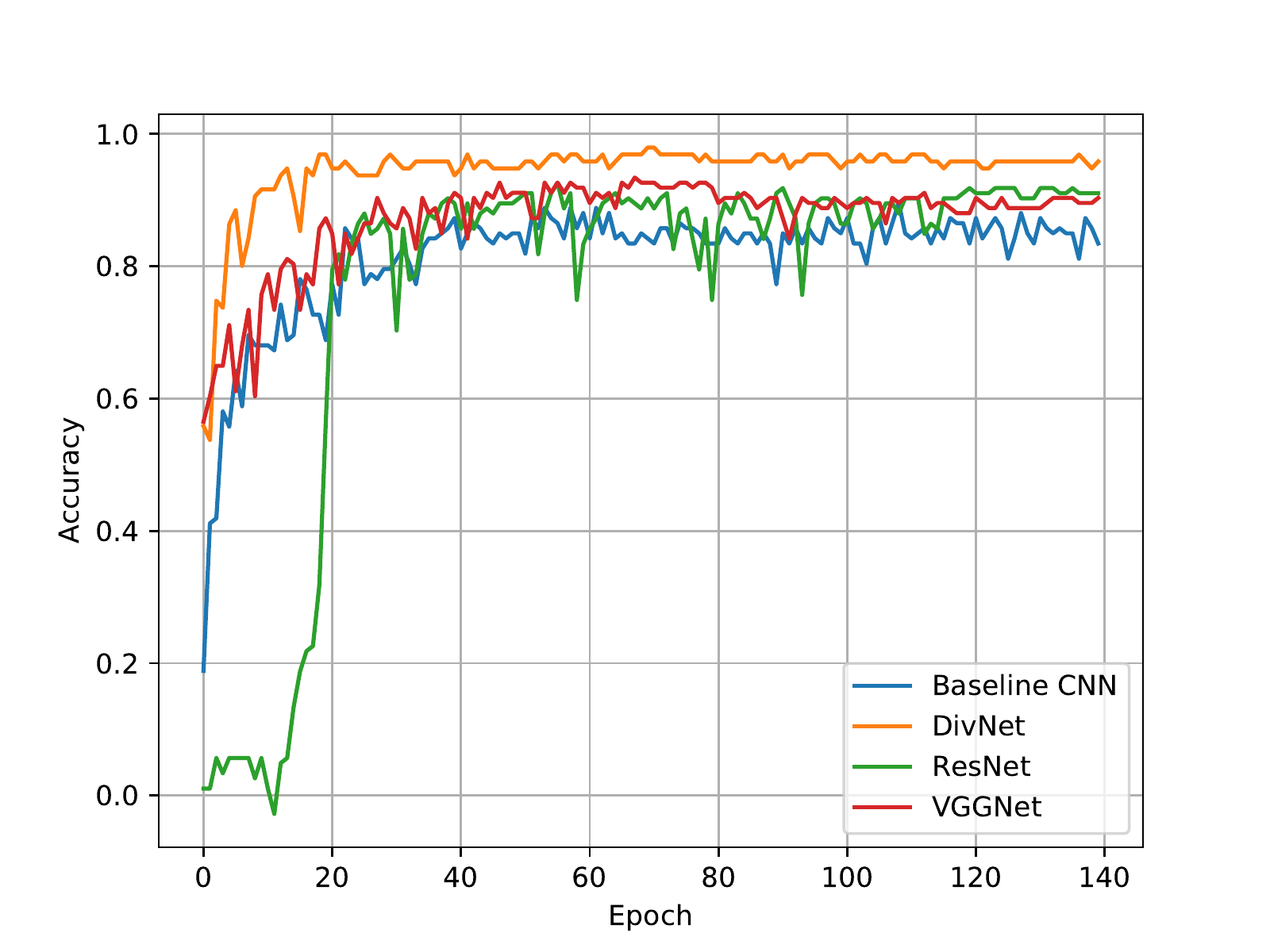}
        \caption{}
	\end{subfigure}
    \caption{Validation accuracy for each method taking $90$x$120$ pixel inputs: (a) bottleneck feature performance for 7-class case, (b) performance after fine-tuning for 7-class case, and (c) performance after fine-tuning for generalization to 11-class case.}
    \label{fig:graphs}
\end{figure*}

The results for validation accuracy with bottleneck features are given in Figure \ref{fig:graphs}(a) for VGGNet, ResNet-50, and the proposed DivNet-15 for the classification of the 7-class human micro-Doppler data. ResNet-50 is found to give roughly a 15\% performance improvement over VGGnet.  This improvement may be attributed to the impact of residual learning and its facilitation of deeper, better optimized neural networks.  However, the proposed DivNet-15 trained on diversified data, which is also a residual network, offers an additional accuracy of 15\% over ResNet-50 and 30\% over VGGnet. This represents a substantial improvement and is a testament to the benefits of pre-training on diversified simulations of micro-Doppler instead of optical imagery.

\begin{table*}[t]
\centering
\caption{Confusion matrix for DivNet with overall test accuracy of 97\% at 40 dB SNR.}
\label{tab:confmat7class}
\begin{tabular}{|
>{\columncolor[HTML]{C0C0C0}}c |c|c|c|c|c|c|c|}
\hline
\textbf{\%}      & \cellcolor[HTML]{C0C0C0}\textbf{Walking}                    & \cellcolor[HTML]{C0C0C0}\textbf{Jogging}                    & \cellcolor[HTML]{C0C0C0}\textbf{Limping}                     & \cellcolor[HTML]{C0C0C0}\textbf{Falling}                    & \cellcolor[HTML]{C0C0C0}\textbf{Sitting}                    & \cellcolor[HTML]{C0C0C0}\textbf{Cane}                        & \cellcolor[HTML]{C0C0C0}\textbf{Walker}                       \\ \hline
\textbf{Walking} & \cellcolor[HTML]{32CB00}{\color[HTML]{333333} \textbf{100}} & 0                                                           & 0                                                            & 0                                                           & 0                                                           & 0                                                            & 0                                                             \\ \hline
\textbf{Jogging} & 0                                                           & \cellcolor[HTML]{32CB00}{\color[HTML]{333333} \textbf{100}} & 0                                                            & 0                                                           & 0                                                           & 0                                                            & 0                                                             \\ \hline
\textbf{Limping} & 0                                                           & 0                                                           & \cellcolor[HTML]{32CB00}{\color[HTML]{333333} \textbf{97.2}} & 0                                                           & 0                                                           & \cellcolor[HTML]{FFFE65}2.8                                                          & 0                                                             \\ \hline
\textbf{Falling} & 0                                                           & 0                                                           & 0                                                            & \cellcolor[HTML]{32CB00}{\color[HTML]{333333} \textbf{100}} & 0                                                           & 0                                                            & 0                                                             \\ \hline
\textbf{Sitting} & 0                                                           & 0                                                           & 0                                                            & 0                                                           & \cellcolor[HTML]{32CB00}{\color[HTML]{333333} \textbf{100}} & 0                                                            & 0                                                             \\ \hline
\textbf{Cane}    & 0                                                           & 0                                                           & 0                                                            & 0                                                           & 0                                                           & \cellcolor[HTML]{32CB00}{\color[HTML]{333333} \textbf{96.4}} & \cellcolor[HTML]{FFFE65} 3.6                                                           \\ \hline
\textbf{Walker}  & 0                                                           & 0                                                           & 0                                                            & 0                                                           & 0                                                           & \cellcolor[HTML]{FD6864}15.41                                & \cellcolor[HTML]{32CB00}{\color[HTML]{333333} \textbf{84.59}} \\ \hline
\end{tabular}
\end{table*}

\subsection{Classification Accuracy and Target Generalization}
The classification accuracy given by using the baseline 4-layer CNN conventionally trained with measured data only, transfer learning from optical imagery, and the proposed DivNet-15 trained on diversified data is compared in Figure \ref{fig:graphs}(b).  The weights of all networks have been fine-tuned with 474 samples of measured data spanning 7 target classes subsequent to initial pre-training.  Notice that this second training stage yields a significant improvement over the initial pre-training offered with optical imagery - increasing accuracy to to 94\% for VGGnet and 95\% for ResNet.  Meanwhile, conventional training on a small set of measured data alone with the 4-layer baseline CNN results in the worst performance with an accuracy of 86\%.  In contrast, the proposed DivNet-15 pre-trained on diversified data outperforms all other methods with 97\% test accuracy.

This result is significant not only in that it substantially reduces the amount of human resources required to procure the training dataset, but also in that it enables high performance even when the test subject has never yet been observed before:  i.e., good target generalization.  The Kinect-based MOCAP measurements are collected at Villanova University on a complete distinct set of test subjects than the test data set, which was collected at TOBB University.  The ability to generalize to previously unobserved targets is an important result when considered that this situation is typical in security applications and that conventional classifiers have been known to exhibit significant losses on the order of tens of percent in accuracy when the subjects comprising the test data differ from that of the training data \cite{padar}.

Finally, a confusion matrix showing the classification performance of DivNet-15 for each activity is given in Table \ref{tab:confmat7class}.  It may be observed that the primary source of confusion occurs between the classes of walking with a cane versus using a walker, while all other classes have been correctly identified with at least 97\% accuracy.  Reasons for this confusion may be the inherent similarity between classes, and that the current simulations of micro-Doppler include only reflections due to the human body, not due to walking aides.

\subsection{Performance Under Noise}

\begin{figure*}[t]
	\begin{subfigure}[t]{.32\linewidth}
		\includegraphics [width=2.5in] {40db7classesfinetuning4methods.pdf}
		\caption{SNR = 40 dB}
	\end{subfigure}
	\begin{subfigure}[t]{.32\linewidth}
		\includegraphics [width=2.5in] {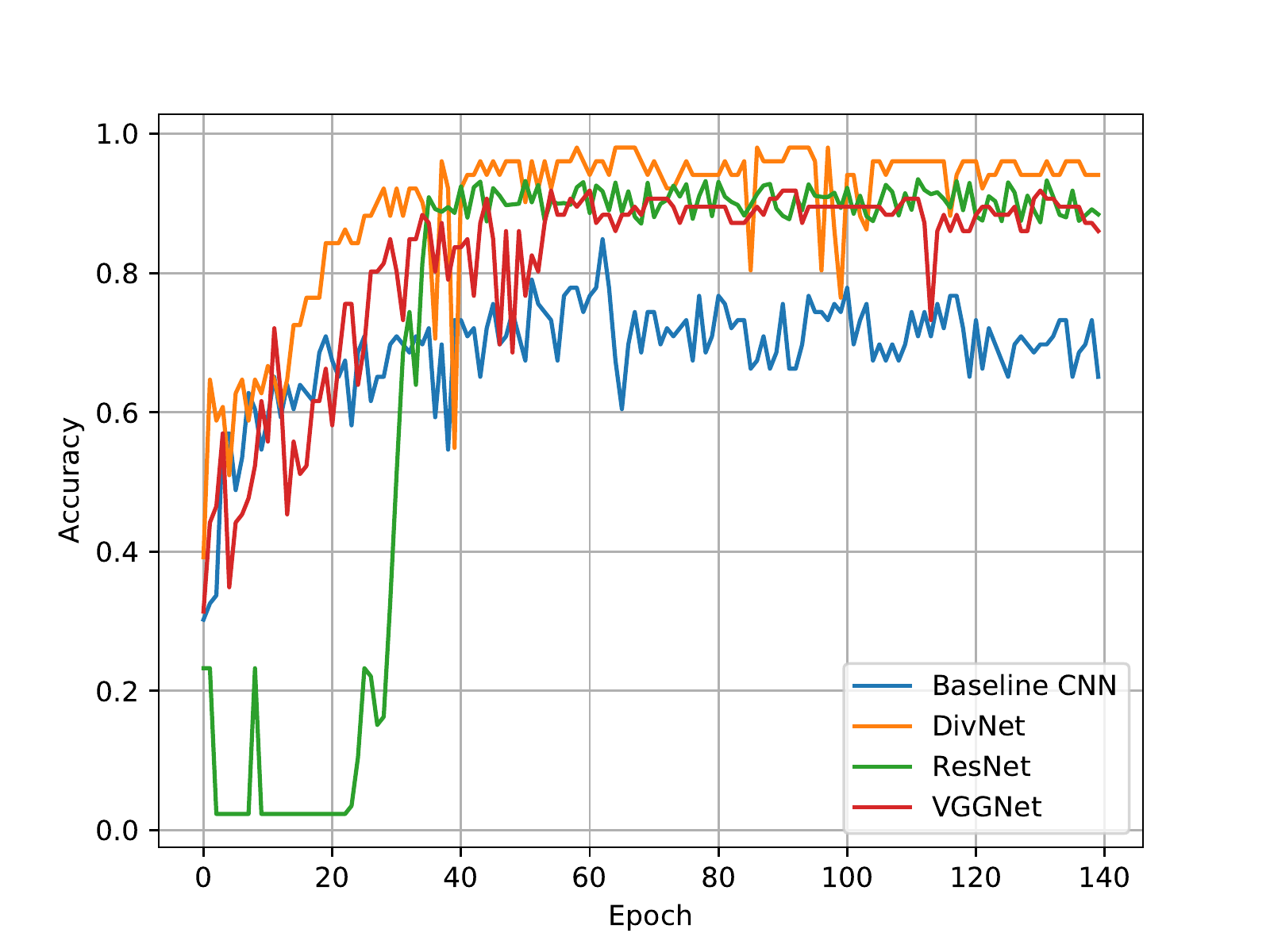}
		\caption{SNR = 30 dB}
	\end{subfigure}
    \begin{subfigure}[t]{.32\linewidth}
		\includegraphics [width=2.5in] {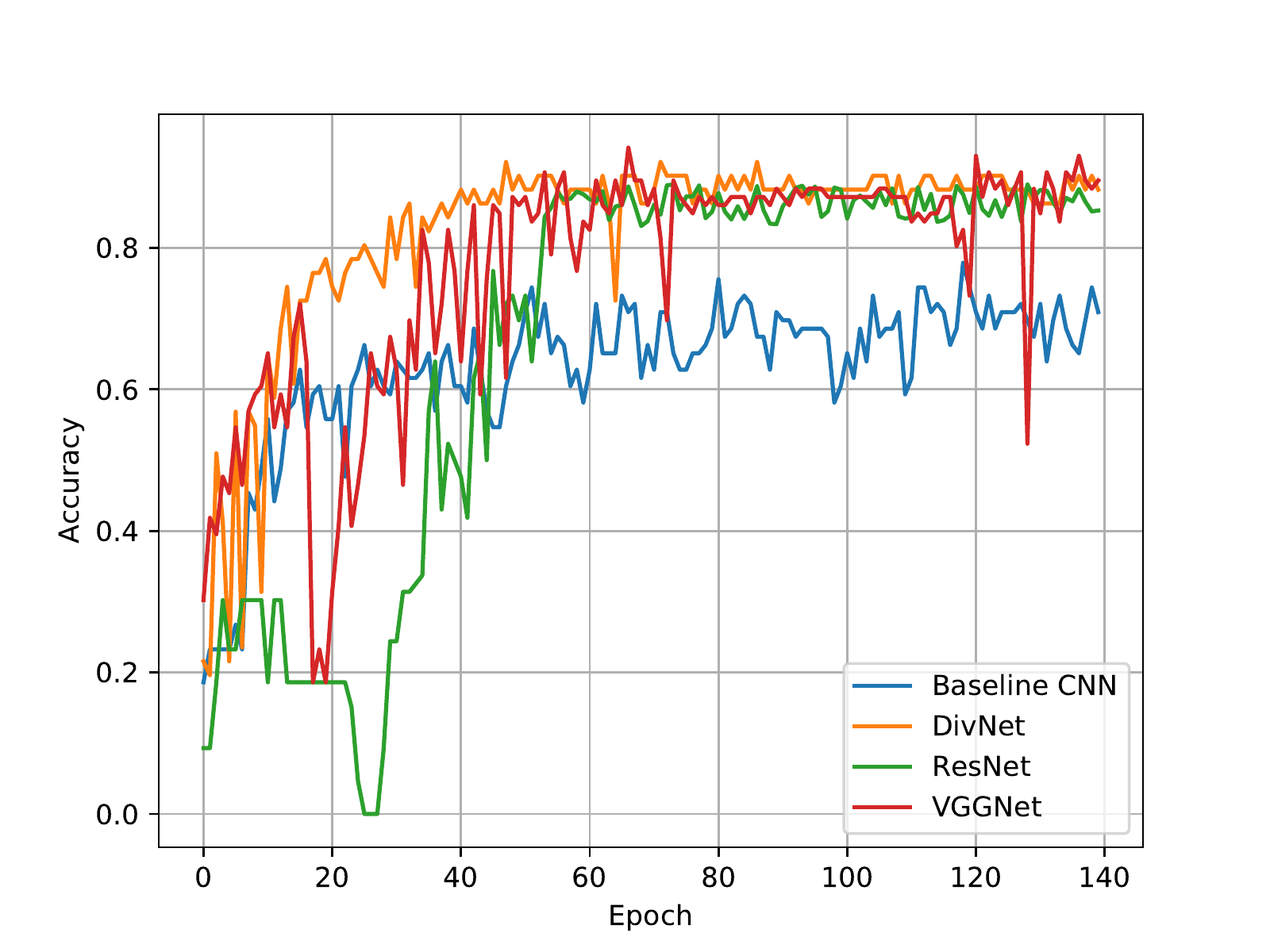}
		\caption{SNR = 15 dB}
	\end{subfigure}
    \caption{Validation accuracy for classification of 7 activities with $90$x$120$ pixel inputs.}
    \label{fig:7classesfinetuningall}
\end{figure*}

First, the classification performance for each method under varying levels of noise is considered for the 7-class problem.  The validation accuracies attained as a function of epoch are shown in Figure \ref{fig:7classesfinetuningall} for three different signal-to-noise ratios (SNRs):  40 dB, 30 dB and 15 dB.  Notice that the proposed DivNet approach exhibits much more consistency during the training process, as well as rapid convergence near about 40 epochs at a validation accuracy that surpasses all other methods at high SNR (30 dB and 40 dB).  When noise is relatively high, such as at 15 dB SNR, all transfer learning methods perform roughly the same, greatly surpassing the baseline CNN by about 15\%. Thus, in general, it may be observed that transfer learning approaches are more robust against noise than CNNs using only real data for training. However, despite being a much deeper network, ResNet-50 offers close to the same performance as VGGNet - an indication of the limitations incurred due to training on a transfer domain substantially different in phenomenology of the target domain (i.e. optical versus radio frequency).  In essence, the proposed method offers the best of both worlds: reaping the benefits of transfer learning from a training set that may be made arbitrarily large, while maintaining the physical relevance and similarities between training and test sets.  Test accuracies achieved at different SNRs for all methods are tabulated in Table \ref{tab:test7class}.  The proposed method surpasses other methods at high SNR.  At low SNR, all transfer learning approaches clearly surpass the performance attained by the baseline CNN.

\begin{table}[t!]
\centering
\caption{7-class test accuracy for all methods.}
\begin{tabular}{|c|c|c|c|}
\hline
\textbf{Method} & \textbf{40 dB} & \textbf{30 dB} & \textbf{15 dB} \\
\hline\hline
\cellcolor{skyblue} Proposed DivNet-15 & \cellcolor{skyblue}97\% & \cellcolor{skyblue}89\% & \cellcolor{skyblue}85\% \\
ResNet-50 & 95\% & 88\% & 85\% \\
VGGnet & 94\% & 87\% & 86\% \\
Baseline CNN & 86\% & 70\% & 69\% \\
\hline
\end{tabular}
\label{tab:test7class}
\end{table}



\subsection{Class Generalization Performance}
Another benefit of the diversified data on training is its potential to improve the initialization of DNNs when classifying a greater number of classes than for which MOCAP-based training data exists.  For example, consider the classification of 11 activities, where in addition to the 7 activities considered before, measured data for creeping, crawling, using a wheelchair and walking with crutches has also been acquired.  However, suppose that MOCAP data for only the original 7-classes has been acquired.  Using the proposed methodology, we will thus initially pre-train on 7-classes of diversified data, while fine-tuning on 11-classes of measured data.  

Table \ref{11classtable} shows the classification performance achieved as a function of DNN depth.  In this case, overfitting is not observed until a depth of 20 convolutional layers.  The maximum accuracy is achieved with 15 layers, as in the 7-class case; hence, in Figure \ref{fig:graphs}(c), we next compare the performance of DivNet-15 to that achieved using VGGnet and ResNet trained on optical imagery, as well as the baseline 4-layer CNN trained on measured data only.  The proposed approach again yields the highest accuracy of 95.7\%, significantly higher than the conventionally trained baseline CNN that gives 87\% accuracy, as well as transfer learning from optical imagery:  VGGnet and ResNet both yield an accuracy of 91\%. 

A confusion matrix showing the classification performance of DivNet-15 for the 11-class activity recognition problem is given in Table \ref{tab:11classconfmat}.  As before, a significant source of confusion is cane versus walker usage.  But now the primary confusion is observed between creeping and crawling.  This is not unexpected due to the similarity between classes, and the tendency for subjects to push-off on the knees when trying to perform a military style creep.  All other classes, however, are correctly identified with at least 96\% accuracy.

These results show that simulated MOCAP data for each class is not necessarily required - if MOCAP data for a given class is for any reason unavailable or not acquirable, pre-training on diversified data from similar classes can still be used to minimize measured data requirements through fine-tuning. 

\begin{table}[!t]
	\centering
	\caption{Classification performance for 11-class case for residual DivNet pre-trained on 7-classes of 90x120 pixel diversified data (overfitting cases highlighted)}
	\begin{tabular}{|c|c|c|c|}
\hline
\textbf{No. of Conv. Layers}&  \textbf{Train Acc.} & \textbf{VA} & \textbf{TA}  \\
 \hline\hline
\textbf{5} 
   & 100\%  & 93\% & 91\%   \\
 \hline
 
 \textbf{10} 
   & 100\%  & 94.5\% & 95\%   \\
 \hline
 
 \textbf{15} 
   & 100\%  & 96\% & 96\%   \\
 \hline
 
 \textbf{20} 
   & \cellcolor{yellow!70} 100\%  & \cellcolor{yellow!70} 90\% & \cellcolor{yellow!70} 91\%   \\
 \hline
 
 \textbf{25} 
   & \cellcolor{yellow!70} 100\%  &  \cellcolor{yellow!70} 80.5\% & \cellcolor{yellow!70} 76\%   \\
 \hline
 
 \textbf{30} 
   & \cellcolor{yellow!70} 100\%  & \cellcolor{yellow!70} 9\% & \cellcolor{yellow!70} 9\%   \\
 \hline
\end{tabular}
\label{11classtable}
\end{table}

\begin{table*}[]
\centering
\caption{Confusion matrix with DivNet-15 for 11-class case with overall test accuracy of 95.7\%.}
\label{tab:11classconfmat}
\begin{tabular}{|
>{\columncolor[HTML]{C0C0C0}}c |c|c|c|c|c|c|c|c|c|c|c|}
\hline
\textbf{\%}         & \cellcolor[HTML]{C0C0C0}\textbf{Walking}           & \cellcolor[HTML]{C0C0C0}\textbf{Jogging}           & \cellcolor[HTML]{C0C0C0}\textbf{Limping}           & \cellcolor[HTML]{C0C0C0}\textbf{Falling}           & \cellcolor[HTML]{C0C0C0}\textbf{Sitting}           & \cellcolor[HTML]{C0C0C0}\textbf{Cane}              & \cellcolor[HTML]{C0C0C0}\textbf{Walker}              & \cellcolor[HTML]{C0C0C0}\textbf{Crutches}          & \cellcolor[HTML]{C0C0C0}\textbf{Wheelchair}       & \cellcolor[HTML]{C0C0C0}\textbf{Crawling}           & \cellcolor[HTML]{C0C0C0}\textbf{Creeping}            \\ \hline
\textbf{Walking}    & \cellcolor[HTML]{32CB00}{\color[HTML]{000000} 100} & 0                                                  & 0                                                  & 0                                                  & 0                                                  & 0                                                  & 0                                                    & 0                                                  & 0                                                 & 0                                                   & 0                                                    \\ \hline
\textbf{Jogging}    & 0                                                  & \cellcolor[HTML]{32CB00}{\color[HTML]{000000} 100} & 0                                                  & 0                                                  & 0                                                  & 0                                                  & 0                                                    & 0                                                  & 0                                                 & 0                                                   & 0                                                    \\ \hline
\textbf{Limping}    & 0                                                  & 0                                                  & \cellcolor[HTML]{32CB00}{\color[HTML]{000000} 100} & 0                                                  & 0                                                  & \cellcolor[HTML]{FFFFFF}0                          & 0                                                    & 0                                                  & 0                                                 & 0                                                   & 0                                                    \\ \hline
\textbf{Falling}    & 0                                                  & 0                                                  & 0                                                  & \cellcolor[HTML]{32CB00}{\color[HTML]{000000} 100} & 0                                                  & 0                                                  & 0                                                    & 0                                                  & 0                                                 & 0                                                   & 0                                                    \\ \hline
\textbf{Sitting}    & 0                                                  & 0                                                  & 0                                                  & 0                                                  & \cellcolor[HTML]{32CB00}{\color[HTML]{000000} 100} & 0                                                  & 0                                                    & 0                                                  & 0                                                 & 0                                                   & 0                                                    \\ \hline
\textbf{Cane}       & 0                                                  & 0                                                  & 0                                                  & 0                                                  & 0                                                  & \cellcolor[HTML]{32CB00}{\color[HTML]{000000} 100} & \cellcolor[HTML]{FFFFFF}0                            & 0                                                  & 0                                                 & 0                                                   & 0                                                    \\ \hline
\textbf{Walker}     & 0                                                  & 0                                                  & 0                                                  & 0                                                  & 0                                                  & \cellcolor[HTML]{FD6864}13.14                      & \cellcolor[HTML]{32CB00}{\color[HTML]{000000} 86.86} & 0                                                  & 0                                                 & 0                                                   & 0                                                    \\ \hline
\textbf{Crutches}   & 0                                                  & 0                                                  & 0                                                  & 0                                                  & 0                                                  & 0                                                  & 0                                                    & \cellcolor[HTML]{32CB00}{\color[HTML]{000000} 100} & 0                                                 & 0                                                   & 0                                                    \\ \hline
\textbf{Wheelchair} & 0                                                  & 0                                                  & \cellcolor[HTML]{FCFF2F}4                          & 0                                                  & 0                                                  & 0                                                  & 0                                                    & 0                                                  & \cellcolor[HTML]{32CB00}{\color[HTML]{000000} 96} & 0                                                   & 0                                                    \\ \hline
\textbf{Crawling}   & 0                                                  & 0                                                  & 0                                                  & 0                                                  & 0                                                  & 0                                                  & 0                                                    & 0                                                  & 0                                                 & \cellcolor[HTML]{32CB00}{\color[HTML]{000000} 88.4} & \cellcolor[HTML]{FD6864}11.6                         \\ \hline
\textbf{Creeping}   & 0                       & 0                                                  & 0                                                  & 0                                                  & 0                                                  & 0                                                  & 0                                                    & 0                                                  & 0                                                 & \cellcolor[HTML]{FD6864}18.47                       & \cellcolor[HTML]{32CB00}{\color[HTML]{000000} 81.53} \\ \hline
\end{tabular}
\end{table*}

\begin{figure*}[t!]
\centering
	\begin{subfigure}[t]{.42\textwidth}
    \centering
		\includegraphics [width=2.65in] {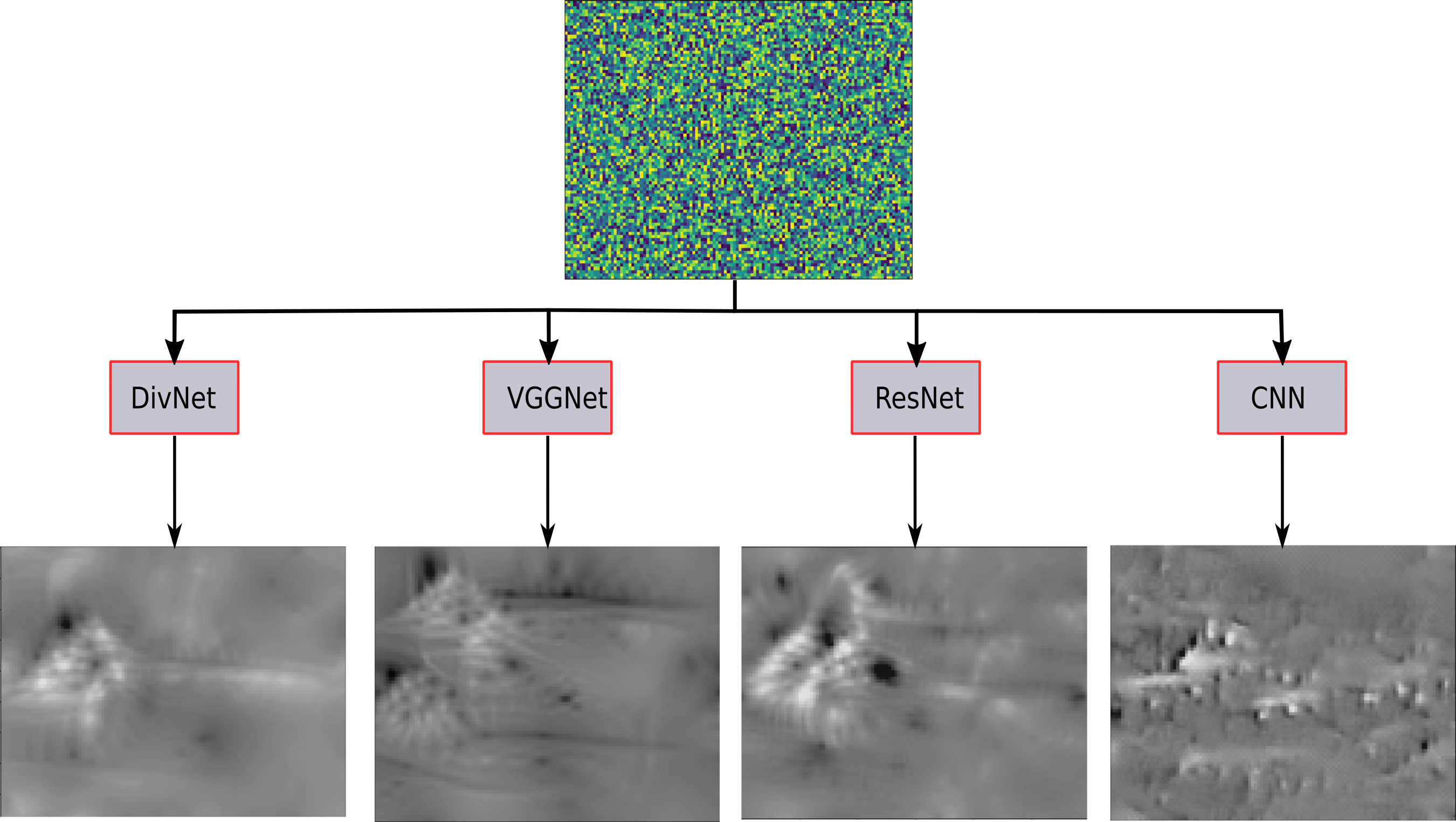}
        \caption{}
	\end{subfigure}
    \begin{subfigure}[t]{.42\textwidth}
    \centering
		\includegraphics [width=2.8in] {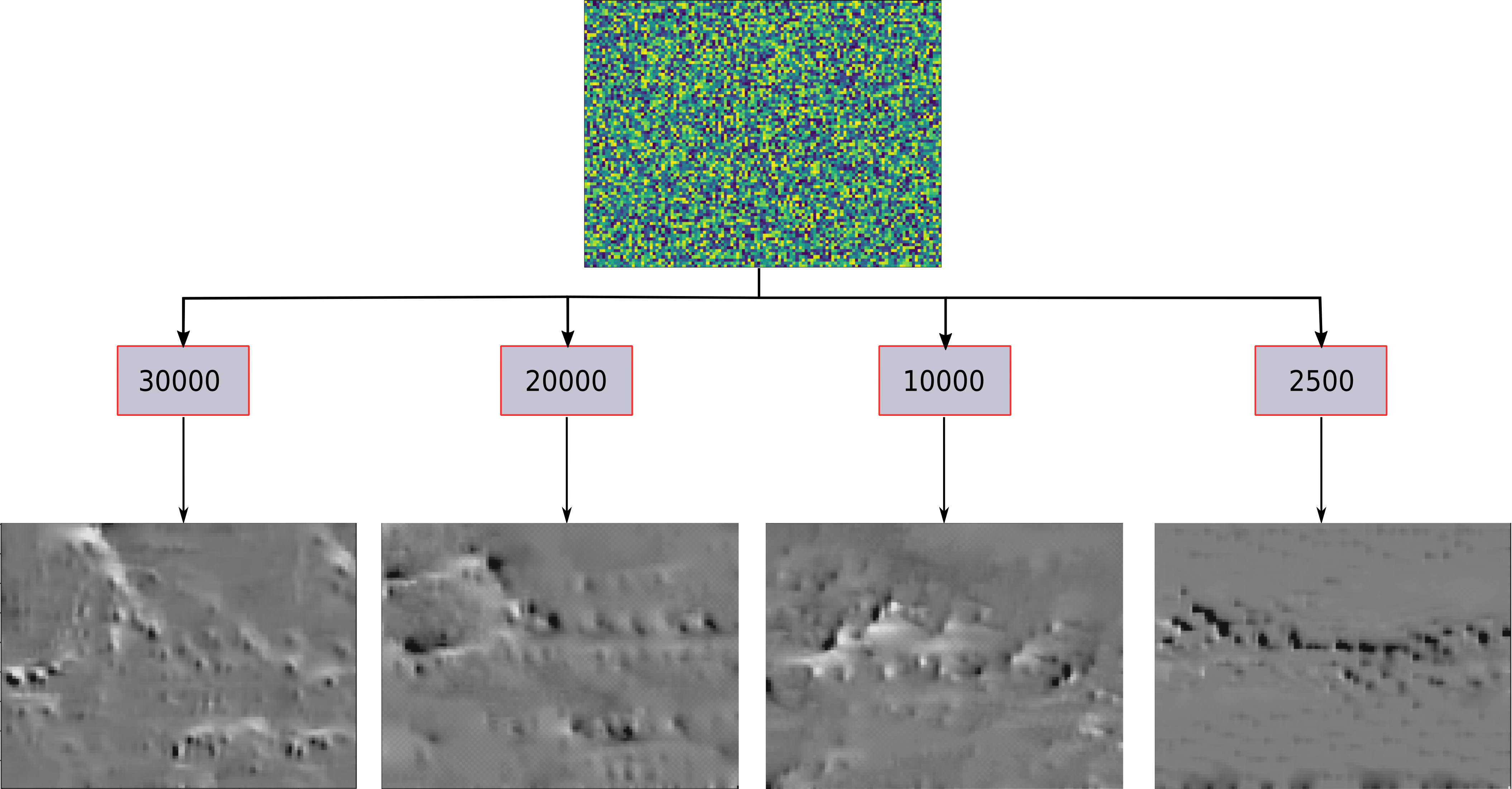}
        \caption{}
	\end{subfigure}
    \caption{Class visualizations: (a) Falling class, after fine tuning with measured data, for each compared method; and (b) falling class for different numbers of diversified samples used in pre-training.}
    \label{fig:vis}
\end{figure*}

\subsection{Class Visualization Comparison}

To better understand the internal workings of the DNNs considered in this work, a technique for visualizing learned class representations is utilized.  The approach is based on reconstructing an image $x$ from its representation $\Phi(x)$ through minimization of the following objective function:
\begin{equation}
x^* = \underset{x}{\operatorname{argmin}}\ R_{{TV}^\beta}(x) + \norm{x}^\alpha_\alpha + -\frac{1}{Z}\langle \Phi(x),\Phi_0\rangle 
\label{eqn:classmaxeq}
\end{equation}
where $x^*$ is the reconstructed image, and $Z$ is a normalization constant.  The term $\norm{x}^\alpha_\alpha$ regularizes the objective function by forcing the intensity of the pixels to stay within a fixed boundary, while the term $R_{{TV}^\beta}(x)$ denotes the total variation of an image $x$:
\begin{dmath}
R_{{TV}^\beta}(x) = \frac{1}{HWV^\beta} \sum\limits_{uvk}\Big( (x(v,u,+1,k) - x(v,u,k))^2 + (x(v+1,u,k) - x(v,u,k))^2 \Big)^{\frac{\beta}{2} }
\end{dmath}
where $H$ and $W$ refers to the number of rows and columns of an image $x$, respectively, and $V$ is the norm of the gradient. Also \textit{u} represents the image columns, \textit{v} represents the image rows and $k$ represents the color channels. The $TV$ regularization term has the effect of penalizing the gradients for large values. It also explicitly forces the representation to have piece-wise constant patches to enable a clean representation \cite{mahendran2016visualizing}. The last term denotes the loss function, which compares the class representation $\Phi(x)$ with the target image $\Phi_0$ using an inner product.  

To visualize what the network thinks best represents a certain class, the network is supplied a randomly generated image, while each class is assigned a score, and the optimization in Eq. \ref{eqn:classmaxeq} done so that the score of other classes is minimized \cite{simonyan2013deep}. In this work, this is done by  using a linear activation function (by replacing the softmax function), and choosing the values of parameters $\beta$ and $V$ as 3 and 20, respectively, while $\alpha$ is selected as 6. The Equation \ref{eqn:classmaxeq} is solved with a gradient based solver for 500 iterations with a learning rate of 0.01 for the class of falling.  The representations obtained are shown in Figure \ref{fig:vis}(a) for each model.  
    
From Figure \ref{fig:vis}(a), it may be observed that the proposed DivNet approach yields the class representation that best matches a spectrogram for falling.  VGGnet also captures the initial burst of energy in the image corresponding to the fall, while neither CNN nor ResNet produce intuitive representations.

We have also investigated how training data size affects the class visualizations. In that regard, Figure \ref{fig:vis}(b) is obtained for falling class by training the DivNet architecture with 2500, 10000, 20000 and 30000 simulated training samples. It can be seen that the learned class visualizations are getting more representative as the number of training data increases. However, even 2500 samples are seen to be producing relevant results.

\section{Implementation Considerations}

\subsection{Computational Complexity}
The proposed DivNet-15 architecture for residual transfer learning from diversified micro-Doppler signatures has significantly less computational complexity in comparison to other transfer learning networks, while offering better performance.  Consider the comparison of computational time and network parameters given in Table \ref{tab:compcomplex}, where all computations are done on a NVIDIA Tesla K80 GPU which has a 24 GB of VRAM.  With the smallest number of convolutional parameters, the baseline CNN is fastest in terms of time per epoch, but worst in terms of accuracy.  Among transfer learning methods, the proposed DivNet architecture not only has the greatest accuracy, but also implements a less complex network (fewer total parameters and convolutional parameters) with fastest time per epoch.  We attribute the performance gains with less complexity to training on signatures that are more closely related to target domain - the network is able to learn more quickly the more relevant features to the underlying classification problem.  


\begin{table}[t!]
\centering
\caption{Computational complexity for all methods.}
\begin{tabular}{|c|c|c|c|c|}
\hline
& \textbf{CNN} & \textbf{VGGnet} & \textbf{ResNet} & \cellcolor{skyblue} \textbf{DivNet} \\
\hline\hline
 \textbf{Time / Epoch} & 0.4 s & 9 s & 20 s & \cellcolor{skyblue} 1 s  \\
 \hline
\textbf{Total Params} & 124,525 & 15,200,000 & 23,866,253 &
\cellcolor{skyblue} 4,814,000 \\
\hline
\textbf{Conv. Params} &  28,064 & 14,700,000 & 23,587,712 &
\cellcolor{skyblue} 1,614,091 \\
\hline
\end{tabular}
\label{tab:compcomplex}
\end{table}

\subsection{Diversified Database Size}
The amount of training data utilized can not only limit the depth of DNNs, but can also affect the overall accuracy of the network when there is no overfitting.  Table \ref{accvsampnum} shows how the validation and test accuracies depend upon the number of diversified samples used in pre-training.  When the size of training sample support is low, such as for the cases of 7500 or fewer, note that there is a significant degradation in test accuracy from the maximum achievable with ample sample support (e.g. 30,000 samples).  For the 7-class and 11-class cases, the difference in having 30,000 samples versus just 2,500 samples is as much as 13\% and 14\%, respectively.  These results also show, however, that once a minimum critical size has been reached, generating an even large training database offers minimal gains.  For example, increasing the training sample support from 15,000 to 30,000 samples results in only a 1\% increase in test accuracy.

\begin{table}[h!]
	\centering
	\caption{Performance vs. training sample support size.}
	\begin{tabular}{|c|c|c||c|c|}
\hline
 & \multicolumn{2}{c||}{\textbf{7-class Case}} 
 & \multicolumn{2}{c|}{\textbf{11-class Case}}
\\
\textbf{Sample \#}&  \textbf{VA@500th epoch} & \textbf{TA} & \textbf{VA@500th epoch} & \textbf{TA} \\
 \hline\hline
\textbf{2500} 
& 84\%  & 84\%   & 84.5\%  & 81\%   \\
 \hline
 
 \textbf{5000} 
& 92\%  & 90\%   & 88\%  & 89\%    \\
 \hline
 
 \textbf{7500} 
& 91\%  & 90\%  & 88\%  & 88\%    \\
 \hline
 
 \textbf{15000} 
 & 96\%  & 95.5\%   & 96\%  & 95\%   \\
 \hline
 
 \textbf{30000} 
 & 98\%  & 97\%   & 97\%  & 96\%   \\
 \hline
 
\end{tabular}
\label{accvsampnum}
\end{table}

\section{Conclusion}
This work proposed a new method for training deep neural networks for radar micro-Doppler classification. This method generates possible variants of human motions by exploiting MOCAP simulations to diversify Kinect-based measurements.  The approach not only allows the generation of unlimited training data, but also enables these data to span a wide range of micro-Doppler signatures that are closely resembling or accurately representing those corresponding to the ground truth. This was done by accounting for variations in body size, speed, and individual gait style.  Residual transfer learning was proposed with initial pre-training on diversified signatures to construct deeper neural networks with increased accuracy.  The proposed approach minimizes the requirement for a large number of measured samples; instead, just 474 measured samples were needed to fine-tune the network and achieve a high classification accuracy of 97\% for a 7-class activity recognition problem.  Additionally, it was shown that the proposed approach yields high target and class generalization performance, outperforming conventional training on small measured data sets and transfer learning from optical imagery.

\ifCLASSOPTIONcaptionsoff
  \newpage
\fi



\bibliographystyle{IEEEtran}
\bibliography{references.bib}

\end{document}